\documentclass[twocolumn]{autart}             

\usepackage{enumerate}
\usepackage{amsfonts,amsmath}
\usepackage{amssymb}
\usepackage{graphicx}          
\usepackage{subeqnarray}

\DeclareMathOperator{\tr}{Tr}

\usepackage{color}
\usepackage[normalem]{ulem}
\usepackage{soul} 
\soulregister\cite7
\soulregister\ref7
\soulregister\pageref7

\begin{document}

\begin{frontmatter}

\title{Ground-state Stabilization of Quantum Finite-level Systems by Dissipation\thanksref{footnoteinfo}\thanksref{funding}} 

\thanks[footnoteinfo]{This paper was not presented at any IFAC
meeting. Corresponding author Y.~Pan. Tel. +02-612-58601.}

\thanks[funding]{This research was supported under Australian Research Council's Discovery Projects funding scheme (Projects DP140101779 and DP110102322).}

\author[Yu]{Yu Pan}\ead{yu.pan.83.yp@gmail.com},    
\author[Valery]{Valery Ugrinovskii}\ead{v.ugrinovskii@gmail.com},               
\author[Matt]{Matthew~R.~James}\ead{matthew.james@anu.edu.au}  

\address[Yu]{Research School of Engineering, Australian National University, Canberra, ACT 0200, Australia}  
\address[Valery]{School of Engineering and Information Technology, University of New South Wales at ADFA, Canberra, ACT 2600, Australia}             
\address[Matt]{ARC Centre for Quantum Computation and Communication Technology, Research School of Engineering, Australian National University,
Canberra, ACT 0200, Australia}        


\begin{abstract}                          
Control by dissipation, or environment engineering, constitutes an
important methodology within quantum coherent control which was proposed to
improve the robustness and scalability of quantum control systems. The
system-environment coupling, often considered to be detrimental to quantum
coherence, also provides the means to steer the system to desired
states. This paper aims to develop the theory for engineering of the
dissipation, based on a ground-state Lyapunov stability analysis of
open quantum systems via a Heisenberg-picture approach. In particular,
Lyapunov stability conditions expressed as operator inequalities allow a
purely algebraic treatment of the environment engineering problem, which
facilitates the integration of quantum components into a large-scale
quantum system and draws an explicit connection to the
classical theory of vector Lyapunov functions and decomposition-aggregation
methods for control of complex systems. This leads to tractable algebraic
conditions concerning the ground-state stability and scalability of quantum
systems. The implications of the results in
relation to dissipative quantum computing and state engineering are also
discussed in this paper.
\end{abstract}

\end{frontmatter}

\section{Introduction}
Control of quantum systems lies at the core of the quantum technology
\cite{Wiseman09,Dong10,Altafini12}, while stability analysis provides the
appropriate tool for the systematic development of quantum control
theory. The stability analysis has been used in several quantum control
synthesis
problems~\cite{mirra2007,James08,PUJ1,Maalouf11,Qi13,Zhang14}. The
applications include measurement-based feedback control and coherent
control for the generation of quantum states as well as the regulation of
system performance. Among all the methods for stability analysis, the
Lyapunov stability approach is the most fundamental, as the energy of a
quantum system is well-defined for most of the physical systems and a
Lyapunov function can be easily constructed
\cite{Kuang08,wang2010,Sayrin12,Ticozzi122,Amini13}. In particular, as we will
demonstrate in this paper, the Lyapunov method provides a means for
engineering the dissipation to be used as coherent control.

Quantum computing often involves the execution of a sequence of unitary
operations on quantum systems. However, the severe decoherence associated
with the quantum systems presents a major obstacle to the scalability of
this approach. For this reason, methods for robust realization of
unitary operations are currently under discussion. The possible plans
include topological quantum computing, adiabatic quantum computing and
dissipative quantum computing. Among these schemes, the adiabatic quantum
computing and dissipative quantum computing have direct relevance to the
stability of quantum systems. For example, in dissipative quantum computing
and state engineering, dissipation is introduced as a resource to
coherently control the system \cite{Verstraete09}. The idea is to consider
open quantum systems, and stabilize their quantum states by engineering the
system-environment interaction. If designed judiciously, the dissipation
will drive the system to a target steady state regardless of the initial
state and external perturbations. This method can be used to generate
highly entangled quantum states, and perform quantum computation by
encoding the computation result to the steady state of the system.  Since
dissipation of energy is the key physical principle behind this method,
this kind of coherent control approach can be referred to as control by
dissipation. Our goal in this paper is to formulate the method of control
by dissipation within the framework of ground-state stability, and then
propose approaches for the synthesis of system-environment dissipative
interactions that rely on Lyapunov methods for stability analysis.

Stability of quantum states has been the focus of many theoretical
studies. Many of them have successfully derived sufficient conditions for
convergence of quantum Markov systems to a steady state
\cite{Spohn76,Frigerio78,Kraus08,Schirmer10,Koga12,Sauer13,Pan14}. In
particular, the stability of quantum states in a
dissipative setting has been considered in
\cite{Ticozzi09,Schirmer10,Ticozzi12,Altafini12}. In these studies, the
target state is often explicitly given and follows a
Schr{\"o}dinger-picture master equation. The dissipative couplings,
compensated by Hamiltonian control, can generate a Markov process that
converges to the target states \cite{Ticozzi09,Ticozzi12}. The implementation of the
system-environment couplings with the practical resources has been
investigated experimentally. Dissipative engineering of several types
of quantum systems has been demonstrated in recent years
\cite{Barreiro11,Krauter11,Kastoryano11,Lin13,Schindler13,Shankar13}.

In this paper, we adopt an alternative path to approach the stability
theory within the Heisenberg picture, where instead of designating
target states explicitly, they are characterized as ground
states a Lyapunov operator, and the stability problem is
  transformed to the problem of stabilization of the ground states of the
  Lyapunov operator. The formalism of Lyapunov stability can thus be
  conveniently introduced to engineer the desired system dissipation within this
framework. This allows to derive tractable sufficient conditions expressed
in terms of operator inequalities, which can be used for the synthesis of
the desired system-environment coupling. Such conditions is the main
contribution of this paper compared to our previous work~\cite{Pan14}. The
general results regarding stability of Lyapunov operators obtained in
\cite{Pan14} do not readily apply to the problem of control by dissipation.

An important advantage of the Heisenberg-picture approach developed here is
that the target state does not need to be given in advance. In addition to
the entangled-state engineering applications in which the Lyapunov operator
is chosen based on the knowledge of the target state, there exists a large
class of applications where the control goals are posed as minimization of
the expectation of an operator while the target state with respect to
which the expectation is taken is not known. For example, the problems
of sequential quantum computation and the quantum satisfiability problem (SAT)
\cite{Bravyi06,Nielsen04} involve operators which play the role of cost
functions. In these problems, the target states which minimize the expectation of the operators are unknown and result from
computation and/or control. Moreover, the target state in these
applications may be not unique. This complicates the analysis based on the
conventional Schr{\"o}dinger-picture approach. Therefore, the
Heisenberg-picture approach extends the applicability of the control by
dissipation.

One of the main contributions of this paper is concerned with the
scalability of the control by dissipation, when this control method is
applied to large quantum systems comprised of multiple interacting
subsystems coupled with the environment. The Heisenberg-picture
Lyapunov approach has an advantage in that the problem can be treated
in a way that resembles the decomposition-aggregation engineering
\cite{Bellman62,Siljak78} for complex classical systems. Namely, a large-scale
quantum system is decomposed into subsystems and an
individual Lyapunov operator is associated with a subsystem. This allows
us to establish conditions, expressed in terms of the subsystems'  Lyapunov
operators, under which the quantum system is guaranteed to converge to its
ground state. Here we note a similarity with the classical connective stability
conditions \cite{Siljak78}, which have proved to be useful in the
synthesis of decentralized controllers for large-scale systems.

A typical methodology for the synthesis of dissipations involves two problems,
the calculation of the stabilizing system-environment couplings and the
implementation of these couplings  using the available physical resources.
For example, it is possible to construct a  coherent optical network to
realize a linear coupling \cite{Nurdin2009}. Therefore, in this paper we
focus on the first problem of calculation of coupling operators that
render the states of the ground energy asymptotically stable. Particularly,
we can apply this method to check the feasibility of the solutions proposed
in \cite{Verstraete09}. It is worth mentioning that the constraints on the
system-environment couplings could be greatly relaxed if Hamiltonian
control is available \cite{Ticozzi09,Ticozzi12}.

The preliminary version of this paper has been accepted for presentation at the American
Control Conference~\cite{PanUJ1a}. Compared to the preliminary conference
version, this paper has been substantially revised and expanded. It
includes a detailed exposition of the background on open quantum systems,
the new material on the scalability of the Lyapunov methods, synthesis of dissipation, the examples and
an application to stabilization of quantum states associated with quantum toric
codes. The paper also includes detailed proofs of all the results and gives
detailed discussions of these results, which were not included
in~\cite{PanUJ1a}.

The paper is organised as follows. In Section \ref{np}, we introduce the notations and the model considered in
this paper. In Section \ref{ls} we present the ground-state stability
analysis of Lyapunov operators. Section \ref{secscal} discusses the
scalability problem, where a large quantum system may be governed by more
than one Lyapunov operators. Section \ref{secapp} concerns with the
synthesis of the dissipation. More explicitly, this section concerns with
the calculation of the correct coherent couplings for the ground-state
stabilization when the Lyapunov operator is given. Conclusion is given in
Section \ref{conc}. The proofs of the results are given in the Appendix in
Section~\ref{appa}.

\section{Notations and preliminaries}\label{np}

\subsection{Open quantum systems}

Consider a Hilbert space $\mathcal{H}$ and define $\mathcal B(\mathcal H)$ as the space of bounded operators on $\mathcal H$. We only consider
finite-dimensional quantum systems throughout this paper. In other words,
$\mathcal H$ is assumed to be finite-dimensional throughout the
paper. Hence all bounded operators in our case are representable as complex
matrices. Let $X\in\mathcal B(\mathcal H)$. $X^T$ denotes the transpose of
$X$ and $X^\dagger$ is the adjoint of $X$. An operator $X$ is called an
observable if $X^\dagger=X$. The
notation $X\ge 0$ ($X\le 0$) means the operator $X$ is a positive
(negative) semidefinite operator. We write $X>0$ if $X$ is positive
definite. Also, we will use the notation $X\succeq
0$ for positive semidefinite operators $X$ whose smallest eigenvalue is
equal to 0.

Given a bounded observable $X\in\mathcal
B(\mathcal H)$ and a trace class operator $\rho$ on $\mathcal H$, $\langle
X\rangle_\rho $ denotes the trace of $X\rho$, $\langle X\rangle_\rho
=\tr{X\rho}$. When $\rho$ is a density state, i.e,, a matrix whose trace is
equal to $1$, then $\langle X\rangle_\rho$ is the mean value of $X$ evaluated at
the density state $\rho$.

Control by dissipation is implemented by coupling the systems to a
collection of environments. To describe evolution of the quantum system
subject to an environment, in addition to the Hilbert
space $\mathcal{H}$, consider the environment on a Fock space $\mathcal{H}_B$
over $L^2 (\mathbb{R}_+ , dt) $ corresponding to Boson field modes. The
system evolution in the Heisenberg picture is captured through the time
evolution of observables (self-adjoint operators) of the system. More
precisely, the observable $X$ evolves as $X(t)=U(t)^\dagger (X\otimes
I)U(t)$, where $U(t)$ is the unitary evolution operator of the
combined system. The dynamical equation for $X(t)$ can be expressed using
the quantum stochastic differential equation~\cite{Parth1992}
\begin{eqnarray}~\label{eq:dynone}
dX(t)&=&(-i[X,H]+\mathfrak{L}(X))dt+\sum_{k=1}^K[L_k^\dag(t),X(t)]dB_k(t)\nonumber\\
&+&\sum_{k=1}^K[X(t),L_k(t)]dB_k^\dag(t),
\end{eqnarray}
with
\begin{equation}~\label{eq:generator}
\mathfrak{L}(X)=\sum_{k=1}^KL_k^\dagger{X}L_k-\frac{1}{2}L_k^\dagger{L_k}X-\frac{1}{2}XL_k^\dagger{L_k}.
\end{equation}
Here $H$ is the Hamiltonian of the system, $L_k$ describes the
coupling between the system and the $k$-th of the total $K$ environment fields, $B_k(\cdot )$ and
$B_k^\dagger (\cdot )$ are the annihilation and creation processes defined
on $\mathcal{H}_B$. Equation (\ref{eq:dynone}) defines a Markov
process. The generator of this Markov process is determined by
\begin{equation}\label{multi1} \mathcal{G}(X)=-i[X,H]+\mathfrak{L}(X).
\end{equation}
In conjunction with the Heisenberg picture dynamics, the evolution of the density state $\rho_t$ in the Schr{\"o}dinger picture is given by
\begin{equation}\label{scheq}
\dot{\rho}_t=-i[H,\rho_t]+\sum_kL_k{\rho_t}L_k^\dag-\frac{1}{2}L_k^\dagger{L_k}\rho_t-\frac{1}{2}\rho_tL_k^\dagger{L_k}.
\end{equation}

\subsection{System decomposition}
In the sequel, we will make use of the decomposition of the Hilbert
spaces $\mathcal H$ into the tensor product of  Hilbert
spaces $\mathcal H=\bigotimes_i \mathcal{H}_i$. Each
$\{\mathcal H_i\}$ can be thought of as a Hilbert space on which subsystem
$i$ is defined. As a simple example, a system consisting of $M$ interacting
two-level systems (qubits), can be defined in terms of the Hilbert space
$\mathcal{H}=\bigotimes_{i=1}^M \mathcal {H}_i$, where
$\mathcal{H}_i=\mathbb{C}^2$, the two-dimensional Euclidean complex space.
This motivates considering the mentioned decomposition of $\mathcal{H}$.

Furthermore, the system observables $W_\lambda,\ \lambda={1,2,...,N}$, which could be associated to a subset of
$\{\mathcal{H}_i\}$, can be defined through a standard embedding. For example, if some $W_\lambda$ is defined on $I\otimes \ldots\otimes \mathcal{H}_i\otimes \ldots\otimes \mathcal{H}_j\otimes \ldots\otimes  I$, then $W_\lambda$ is said to be associated to $\mathcal{H}_i$ and $\mathcal{H}_j$, or associated to subsystems $i$ and $j$. The coupling operators $\{L_k\},\ k=1,2,...,K$ are defined on $\mathcal{H}$ as well. We say $L_k$ is associated to $W_\lambda$ if $[L_k,W_\lambda]\neq0$.



\subsection{Lyapunov operators and ground state stability of complex
  quantum systems}

We recall the definition of the Lyapunov operator \cite{Pan14}:
\begin{defn}\label{Lyap.def}
A quantum Lyapunov operator $V$ is an observable (a self-adjoint operator)
on a Hilbert space $\mathcal H$ for which the following properties hold:
\begin{enumerate}[(i)]
\item  $V\succeq 0$.
\item $\mathcal{G}(V)\leq 0$.
\end{enumerate}
\end{defn}
One natural choice of the Lyapunov operator is the energy operator of
the system. For example, the Lyapunov operator can be defined by offsetting
a system Hamiltonian $H$ as $V=H-d\succeq0$, where $d$ is the smallest
eigenvalue of $H$.\footnote{In accordance with the common convention of
  quantum physics, the
  identity operator is omitted here and elsewhere, i.e., $H-d$ should be
  understood as $H-dI$.}

In this paper, we restrict our attention to considering the observables
which satisfy the condition
\begin{equation}\label{coun}
[X,H]=0.
\end{equation}
This condition holds, for example, in the state engineering by dissipation
problems
concerned with preparation of the ground states of Hamiltonians, when no
additional Hamiltonian control is used
\cite{Ticozzi12,Verstraete09,Perez08}. In these problems
the observable of interest is $X=H-d$ and (\ref{coun}) holds trivially.
More generally, under condition (\ref{coun}) the evolution of the
observable $X$ described by (\ref{eq:dynone}) is due to the environment.
This allows us to focus entirely on the analysis and synthesis of the
effects associated with the environment, which is the main  objective of this paper.

For any observable $X$ which satisfies (\ref{coun}), the expression
(\ref{multi1}) for the system generator is simplified into
\begin{equation}\label{gnu}
\mathcal{G}(X)=\sum_kL_k^\dagger{X}L_k-\frac{1}{2}L_k^\dagger{L_k}X-\frac{1}{2}XL_k^\dagger{L_k}.
\end{equation}

We use $\rho_0$ to denote the initial density state of the system, and
$\rho_t$ to denote the system state at time $t$. Following Meyer
\cite{Meyer1995}, any convergence of a trajectory in the form of
$\rho_t\rightarrow\rho_\infty$ should be understood as convergence of
probability distributions, i.e., $\rho_t\rightarrow\rho_\infty$ means
$\tr{\rho_t X}\to \tr{\rho_\infty X}$ for all bounded $X$. This is a nice
convergence property which can be studied using the tools of
quantum probability such as the Prokhorov theorem and
tightness~\cite{Meyer1995}. Recall that a sequence $(\rho_n)_{n\geq1}$ of
trace-class operators
on a Hilbert space is tight if for any $\epsilon>0$, there exists a finite
rank projection $P$ and $n_0>0$ such that $\tr{\rho_nP}>1-\epsilon$ for
all $n\geq n_0$. In addition, the Prokhorov theorem states that
$(\rho_n)_{n\geq1}$ is sequentially compact if $(\rho_n)_{n\geq1}$ is
tight. In this paper, since $\mathcal H$ is finite-dimensional, all projections are finite
rank. Therefore, any trajectory $\rho_t$ is tight \cite{Meyer1995,Pan14},
and hence admits a subsequence converging to a limit point $\rho'$.

The ground-state stability of an operator $X$ is defined using the mean of the operator:
\begin{defn}\label{defgs}
Suppose the smallest eigenvalue of an observable $X$ is $d$. $X$ is said to
be asymptotically ground-state stable if
\begin{equation}\label{defigss}
\langle X(t)\rangle_{\rho_0}=\langle X\rangle_{\rho_t}\rightarrow d, \quad
\mbox{as $t\to\infty$}.
\end{equation}
Here $\langle X\rangle_{\rho_t}$ is an alternative representation of $\langle X(t)\rangle_{\rho_0}$ in terms of $\rho_t$.
\end{defn}
Consequently, a Lyapunov operator $V$ is asymptotically ground-state stable
if
\begin{equation}\label{gsdef1}
\langle V(t)\rangle_{\rho_0}=\langle V\rangle_{\rho_t}\rightarrow 0.
\end{equation}

\begin{rem}
\rm We note a parallel between the ground-state stability and convergence of
probability distributions of classical Markov stochastic
processes to stationary distributions.
Precisely, let $x(t)$ be an $R^l$ valued Markov stochastic
process with the transition probability function
$P(t,x,Y)=\mathbf{Prob}(x(t)\in Y|x(0)=x)$, here $Y$ is a measurable
set. It is known that under certain conditions~\cite[Theorem 4.3 and
p.~121]{Khas2}, there exists a unique stationary probability distribution
$\mu(\cdot)$ for $x(t)$ such that for any bounded function $f(x)$, the
expectation of $f(x)$ with respect to $P(t,x,\cdot)$ converges to that with
respect to $\mu$:
\[
\lim_{t\to\infty}\int_{R^l} f(y)P(t,x,dy) =
     \int_{R^l} f(y)\mu(dy).
\]
This property reads that $\mu$ is an attracting equilibrium distribution
for $\{P(t,x,\cdot),t\ge 0\}$. It is analogous to the property $\langle
X\rangle_{\rho_t}\to d=\langle X\rangle_{\rho'}$ considered above.

If, in addition, $Y_0$ is a set with the property
$P(t,x,Y_0)\to 0$, then letting $f(x)$ be the indicator function of $Y_0$,
$f(x)=\chi_{Y_0}(x)$, leads to $\lim_{t\to\infty}\int_{R^l} \chi_{Y_0}(y)P(t,x,dy) =
     \int_{R^l} \chi_{Y_0}(y)\mu(dy)=0$. This property is analogous to (\ref{gsdef1}); it further illustrates the
analogy between stationary equilibrium distributions of Markov processes
and the ground states of asymptotically ground-state stable Lyapunov
observables.
\end{rem}

\begin{defn}
The state trajectory $\rho_t$ is said to converge to a set $S$ if the
  limit points of $\rho_t$ are all contained in $S$.
\end{defn}
This definition is often used to characterize the convergence to an
invariant set in the invariance principle
\cite{mirra2007,wang2010,Pan14}. Denote $Z_X=\{\rho:\langle
X\rangle_\rho=d\}$ to be the set of the ground states of $X$.
\begin{prop}\label{df3}
The state trajectory $\rho_t$ is converging to $Z_X$ if and only if $X$ is asymptotically ground-state stable.
\end{prop}

The proof of this and subsequent results are given in the Appendix in
Section~\ref{appa}.

We will also exploit the notion of dissipation functional:
\begin{defn}(\cite{Lind1976,Frigerio78})
The dissipation functional of an operator $X$ is defined as
\begin{equation}
\mathfrak{D}(X)=\mathcal{G}(X^\dagger X)-\mathcal{G}(X^\dagger)X-X^\dagger\mathcal{G}(X).\label{dissip.functional}
\end{equation}
\end{defn}
The dissipation functional characterizes the dissipation of energy. With a single coupling operator $L$, the
dissipation functional is calculated to be
\begin{equation}\label{dissipator}
\mathfrak{D}(X)=[L^\dagger,X^\dag][X,L],
\end{equation}
and hence $\mathfrak{D}(X)\geq0$.

The first objective of this paper is to develop the ground-state stability
theory using the notion of Lyapunov operator. We will consider Lyapunov
operators which are associated either with a single
quantum system, or a subsystem of the total system. The need in such theory
can be illustrated by the following result from \cite{Pan14}.
\begin{prop}\label{theorem3}
\cite[Theorem 8]{Pan14} Suppose $V$ is a Lyapunov operator of the
system. The state trajectory $\rho_t$ will converge to
  $Z_V$ if $\langle\mathfrak{D}(V)\rangle_\rho>0$ for $\rho\notin Z_V$ and $[\mathcal G(V),V]=0$.
\end{prop}
As one may see, the conditions in Proposition \ref{theorem3} are not fully
algebraic and also they are not easy to verify; the validation of these
condition requires computing the mean $\langle\mathfrak{D}(V)\rangle_\rho$,
which in tern requires $\rho$ to be explicitly given or known. As
discussed, this may not be possible in certain applications. This motivates
us to revisit the ground-state stability
theory for Lyapunov operators, in order to derive ground-state stability
conditions expressed purely in terms of operator inequalities.

The second objective of this paper is to apply this theory to stabilization
of large-scale quantum systems. We consider an operator $W=\sum_{\lambda=1}^N
W_\lambda,\ W_\lambda\succeq0$, as a candidate for the Lyapunov operator of a
large-scale system, where each $W_\lambda$ is an observable associated with one of the subsystems or a group of the subsystems of the system. Such operator sum representations naturally
arise in many problems of control by dissipation, including the preparation of multipartite entangled
states, hence the theory developed in this paper is aimed at these
applications. For more information about the applications of control by
dissipation please refer to
\cite{Verstraete09,Perez08,Ticozzi12}.

In general, the fact that the individual observables
$W_\lambda$ have zero eigenvalue (as implied by the notation
$W_\lambda\succeq0$ $\forall \lambda$) does not guarantee that $W=\sum_{\lambda=1}^N
W_\lambda$ has a zero eigenvalue; in fact $W$
can be positive definite. In the light of the definition  of
the Lyapunov operator, this means that the operator sum $W=\sum_{\lambda=1}^N
W_\lambda$ may result in an operator which does not satisfy formally all
the properties of Definition~\ref{Lyap.def} (recall that by definition,
Lyapunov operators have a zero eigenvalue). For this reason, we use the
notation $W,W_\lambda$ instead of $V$ in Section \ref{secscal} because there is a
possibility that $W,W_\lambda$ cannot be made Lyapunov operators even if they are asymptotically ground-state stable. Similarly, for $W$ to satisfy the condition
$\mathcal{G}(W)\leq0$, the condition $\mathcal{G}(W_\lambda)\leq0$ does not
have to be satisfied for all $W_\lambda$; that is, $W_\lambda$ may not be a
Lyapunov operator either (for $W_\lambda$ to be a Lyapunov operator, the
condition $\mathcal{G}(W_\lambda)\leq0$ must be satisfied).  We will show
in Section \ref{secscal} that the approaches to the engineering of the
ground-state stability of $W$ can be quite different depending on whether or not
$W_\lambda$ can be taken to be a Lyapunov operator of the subsystems.

It is worth mentioning that the issues discussed above are similar to those
arising within the vector
Lyapunov function approach \cite{Bellman62}. In certain situations arising
in the classical stability theory for large-scale systems, it is
more convenient to use a vector Lyapunov function rather than a scalar
function for a large-scale system \cite{Siljak78}. Indeed, in general,
scalar functions comprising the vector Lyapunov function of a stable
large-scale system do not need to be Lyapunov functions
individually. Particularly, the
decomposition-aggregation method used in \cite{Siljak78} to simplify the
analysis by decomposing the large system into several subsystems made
extensive use of the vector Lyapunov function machinery.  When
the subsystems are coupled together, a connective stability condition
will ensure the total system is stable after the aggregation.
In our case, $W$ is the quantum
counterpart of the vector Lyapunov function, and $W_\lambda$ is the quantum
counterpart of the scalar component of that function. Also, the scalability
property discussed in this paper where the ground-state stability of the
operator $W$ is derived from the ground-state stability properties of the
addends $\{W_\lambda\}$, is parallel to the classical decomposition-aggregation
approach mentioned above. Since each $W_\lambda$ may act on several
subsystems and some coupling may affect more than one subsystems, a scalability condition is needed to ensure the
cross-couplings do not undermine the stability of $W$.

The following lemma summarizes the approaches to the stability of a
large-scale quantum system.
\begin{lem}\label{ff}
 Given a collection of observables $W_\lambda\succeq0,\ \lambda=1,2,...,N$, consider $W=\sum_{\lambda=1}^N
  W_\lambda$, whose smallest eigenvalue is $d$.
\begin{enumerate}[(i)]
\item Suppose $\langle W_\lambda\rangle_{\rho_t}\rightarrow0$ for each $\lambda$. Then $\langle
  W\rangle_{\rho_t}\rightarrow0$ and $d=0$.\label{le1}

\item Conversely, suppose $\langle W\rangle_{\rho_t}\rightarrow0$. Then $d=0$ and each $W_\lambda$ is
  asymptotically ground-state stable.\label{le2}
\end{enumerate}
\end{lem}
The proof of the lemma is deferred to the Appendix.

\begin{rem}
\rm If $\langle W\rangle_{\rho_t}\rightarrow d>0$, each $W_\lambda$ is not
necessarily asymptotically ground-state stable. As a simple example to
illustrate this, define $W=W_1+W_2$ with $W_1=\left(\begin{array}{cc}
1&0\\
0&0
\end{array}\right)\succeq0$, $W_2=\left(\begin{array}{cc}
0&0\\
0&1
\end{array}\right)\succeq0$. The smallest eigenvalue of $W$ is $1$. Letting
$\rho_t=\left(\begin{array}{cc}
\rho_t^{11}&\rho_t^{12}\\
{\rho_t^{12}}^*&\rho_t^{22}
\end{array}\right)$
we obtain that the condition $\langle W\rangle_{\rho_t}\to 1$ is equivalent
to  $\rho_t^{11}+\rho_t^{22}\to 1$. Now suppose both $W_1$ and
$W_2$ converge to their ground states, then we must have $\langle W_1\rangle_{\rho_t}=\rho_t^{11}\to 0$
and $\langle W_2\rangle_{\rho_t}=\rho_t^{22}\to 0$ simultaneously, thus
$\rho_t^{11}+\rho_t^{22}\to 0$, which contradicts the condition $\langle W\rangle_{\rho_t}\to 1$; also see statement (i) of
Lemma~\ref{ff}. The established contradiction shows that
$W_1$ and $W_2$ cannot converge to their ground states simultaneously, and
at least one of them is not ground-state stable.
\end{rem}

Lemma \ref{ff} suggests two different approaches to engineering of the
ground-state stability of $W$ (as we mentioned, investigation of such
approaches is
the main objective of this paper), namely, through engineering the
ground-state stability of every $W_\lambda$, or the ground-state stability
of $\sum_{\lambda\in\Lambda^{'}}W_\lambda$. Here
$\Lambda^{'}$ is a subset of the set of all $\lambda$, $\Lambda$. The
trivial case where $\Lambda$ is divided into $\{\Lambda,\varnothing\}$
means that we engineer the ground-state stability of $W$ directly. In this paper by engineering we mean the synthesis of coupling operators between the environment and the system.

We would like to mention that Lemma \ref{ff} has a connection with the notion of frustration-free
Hamiltonian \cite{Verstraete09}. A Hamiltonian $H$ in the form of $H=\sum_\lambda H_\lambda$ is called frustration-free if the ground states of $H$ are also the ground states of every $H_\lambda$. Suppose $H_\lambda\succeq0$, $d$ is the smallest eigenvalue of $H$ and $\rho_g$ is one of the ground states of $H$. Then if $d=0$, we have $\langle H\rangle_{\rho_g}=\sum_\lambda \langle H_\lambda\rangle_{\rho_g}=0$ and so $\langle H_\lambda\rangle_{\rho_g}=0$, which proves that $H$ is frustration-free. Therefore, if $W$ and $W_\lambda$ denote Hamiltonians in Lemma \ref{ff}, the condition $\langle
W_\lambda\rangle_{\rho_t}\rightarrow0$ of (i) and the condition $\langle W\rangle_{\rho_t}\rightarrow0$ of (ii) in fact imply the frustration-freeness of $W$ through proving $d=0$.

The property of the system observables to maintain their smallest
eigenvalue to be $d=0$ while adding the subsystem
Hamiltonians $H_\lambda$ and associated observables
$W_\lambda$ means that the system size can be increased without perturbing
the ground energy. Such a scalability property is often
desired in quantum engineering.

\section{Lyapunov stability of the ground states}\label{ls}
In this section, we consider the generator with one dissipation channel
\begin{equation}\label{eq:Lindblad}
\mathcal{G}(X)=L^\dagger{X}L-\frac{1}{2}L^\dagger{L}X-\frac{1}{2}XL^\dagger{L}.
\end{equation}

Recall that a state $\rho_I$ is an invariant state of the quantum system,
if it satisfies the condition $\langle X(t)\rangle_{\rho_I}=\langle
X\rangle_{\rho_I}$ for any operator $X$ \cite{Frigerio78,Pan14}. Thus we
have $\langle\mathcal{G}(X)\rangle_{\rho_I}=0$ for an invariant state
$\rho_I$.

The next statement gives the quantum version of the Lyapunov's second
method for stability.
\begin{lem}\label{theorem1}
Suppose $V$ is a Lyapunov operator of the system. If $\langle\mathcal{G}(V)\rangle_\rho<0$ for any $\rho\notin Z_V$, then $V$ is asymptotically ground-state stable.
\end{lem}

The proof of the lemma is deferred to the Appendix.
A special case of Lemma \ref{theorem1} is concerned with the generator
satisfying the condition
\begin{equation}\label{gvcv}
\mathcal G(V)\leq-cV,\quad c>0.
\end{equation}
In this case, we can integrate (\ref{gvcv}) to obtain $\langle V(t)\rangle_{\rho_0}\leq e^{-ct}\langle V\rangle_{\rho_0}$. The system exponentially converges to the ground states of $V$.

The exponential convergence condition (\ref{gvcv}) does not describe all the
dynamics that lead to the asymptotical stability of the ground states. Not
all physical systems are exponentially stable. A more general treatment
will involve dealing with the condition $\mathcal G(V)\leq0$. To this end we will make use of the dissipation functional $\mathfrak{D}(V)$.
\begin{lem}\label{theorem4}
If $V$ is a Lyapunov operator of the system satisfying $\mathfrak{D}(V)\geq cV^2$ for some $c>0$, then $V$ is asymptotically ground-state stable.
\end{lem}
Particularly, we can make use of Lemma \ref{theorem4} to obtain the
following result.
\begin{lem}\label{theorem5}
If $V$ is a Lyapunov operator of the system satisfying
$cV\leq\mathfrak{D}(V)$ for some $c>0$, then the state trajectory $\rho_t$ will converge to $Z_V$.
\end{lem}
With multiple dissipation channels, the generator of the Lyapunov operator $V$ is expressed as (\ref{gnu}) and the dissipation functional becomes
\begin{equation}\label{dissimul}
\mathfrak{D}(V)=\sum_{k}[L_k^\dagger,V][V,L_k].
\end{equation}
All the above stability results can be routinely extended to the
multi-channel case.

Now we define two conditions on an observable $X$ as:
\medskip

\noindent\textbf{Condition ES }
An observable $X$ is said to satisfy condition ES, if there exists a
constant $c>0$ such that
\begin{equation}\label{gleq1}
\mathcal G(X)\leq-cX,\quad c>0.
\end{equation}

\noindent\textbf{Condition DS }
An observable $X$ is said to satisfy condition DS, if there exists a
constant $c>0$ such that
\begin{subeqnarray}
&\mathcal G(X)\leq0,&\slabel{gleq2}\\
&cX\leq\mathfrak{D}(X),&c>0.\slabel{gleq3s}\label{gleq3}
\end{subeqnarray}

Using these conditions, the sufficient conditions for the
convergence to the ground states of general quantum systems can be
expressed by operator inequalities, as summarized in the following theorem:
\begin{thm}\label{thl}
If $V$ is a Lyapunov operator of the system satisfying either Condition ES or Condition DS, then the state trajectory $\rho_t$ will converge to $Z_V$.
\end{thm}

It is worth mentioning that under Condition ES, Theorem~\ref{thl}
guarantees that the state trajectory
converges exponentially, hence the name ES - exponential
stability. Likewise, under Condition DS Theorem~\ref{thl} guarantees that
the convergence is asymptotic and
is ensured by dissipation properties of the system, hence the notation DS-
`dissipative stability'.

Below are two examples to illustrate the use of Conditions ES and DS.
\begin{exmp}\label{exam1}\em
Consider a two-level quantum system with the system Hamiltonian
$H=\left(\begin{array}{cc} \frac{1}{2}&0\\
    0&-\frac{1}{2} \end{array}\right)$.
In this example, we wish to engineer a dissipative coupling to govern the
system to the ground state of the operator $V$ which is defined as
\begin{equation}\label{exame10}
V=\left(\begin{array}{cc}
1&0\\
0&0
\end{array}\right).
\end{equation}
Clearly, $V=H+1/2$ and we have $[V,H]=0$ as required.

To verify condition DS, consider a general coupling operator $L$ with
complex entries
$ L=\left(\begin{array}{cc}
l_{00}&l_{01}\\
l_{10}&l_{11}
\end{array}\right)$
and substitute this expression in (\ref{gleq3}). Condition (\ref{gleq2})
reduces to the inequality
\begin{equation}\label{exame3}
\left(\begin{array}{cc}
-|l_{10}|^2& \frac{1}{2}l_{00}^*l_{01}-\frac{1}{2}l_{10}^*l_{11}\\
\frac{1}{2}l_{01}^*l_{00}-\frac{1}{2}l_{11}^*l_{10}& |l_{01}|^2
\end{array}\right)\leq0,
\end{equation}
which can only hold when $l_{01}=0$ and $|l_{10}|^2|l_{11}|^2\le
0$. Also, (\ref{gleq3s}) reduces to the inequality
\begin{equation}\label{exame2}
[L^\dagger,V][V,L]=\left(\begin{array}{cc}
|l_{10}|^2&0\\
0&|l_{01}|^2
\end{array}\right)\geq\left(\begin{array}{cc}
c&0\\
0&0
\end{array}\right).
\end{equation}
Thus,  to satisfy Condition DS we must select $l_{01}=0$, $l_{11}=0$ and
$l_{10}\neq0$. As a result, according to Theorem~\ref{thl}, coupling the
system with the environment using any $L$ of the form of
\begin{equation}\label{exame4}
\left(\begin{array}{cc}
l_{00}&0\\
l_{10}&0
\end{array}\right),\quad l_{10}\neq0,
\end{equation}
will ensure that the state trajectory $\rho_t$ converges to the set $Z_V$,
which in this example consists of only one element
$\rho'=\left(\begin{array}{cc}
0&0\\
0&1
\end{array}\right)$, which is the unique ground state of the operator $V$.

We now confirm this funding by directly obtaining the equilibrium state of
the Schr{\"o}dinger-picture evolution equation~(\ref{scheq}) under the
dissipation control associated with the coupling operator $L$ of the
form~(\ref{exame4}) and proving
its stability. Under $L$ of the form~(\ref{exame4}), the
Schr{\"o}dinger-picture equation~(\ref{scheq}) becomes
\begin{eqnarray*}
\dot{\rho}&=&-\mbox{i}[H,\rho]+L\rho L^\dag-\frac{1}{2}L^\dag L\rho-\frac{1}{2}\rho L^\dag L\\
&=&-\mbox{i}(\left(\begin{array}{cc}
\frac{1}{2}\rho_{00}&\frac{1}{2}\rho_{01}\\
-\frac{1}{2}\rho_{10}&-\frac{1}{2}\rho_{11}
\end{array}\right)-\left(\begin{array}{cc}
\frac{1}{2}\rho_{00}&-\frac{1}{2}\rho_{01}\\
\frac{1}{2}\rho_{00}&-\frac{1}{2}\rho_{11}
\end{array}\right))\\
&&+\left(\begin{array}{cc}
-|l_{10}|^2\rho_{00}&l_{00}\rho_{00}l_{10}^*-\frac{1}{2}\rho_{01}g\\
l_{10}\rho_{00}l_{00}^*-\frac{1}{2}\rho_{10}g&|l_{10}|^2\rho_{00}
\end{array}\right)\\
&=&\left(\begin{array}{cc}
-|l_{10}|^2\rho_{00}&-\mbox{i}\rho_{01}+l_{00}\rho_{00}l_{10}^*-\frac{1}{2}\rho_{01}g\\
\mbox{i}\rho_{10}+l_{10}\rho_{00}l_{00}^*-\frac{1}{2}\rho_{10}g&|l_{10}|^2\rho_{00}
\end{array}\right)
\end{eqnarray*}
with $g=|l_{00}|^2+|l_{01}|^2$. Its equilibrium state $\rho'$ must satisfy the
condition $\dot\rho'=0$, which implies $\rho_{00}'=\rho_{01}'=0$ since $l_{10}\neq0$. As a result, $\rho_{10}'=0$ as well. Therefore, $\rho'=\left(\begin{array}{cc}
0&0\\
0&1
\end{array}\right)$ is the unique equilibrium state under the dissipation
control of the form~(\ref{exame4}). It also follows from the above
calculation that
\begin{eqnarray*}
  &&\rho_{00,t}=e^{-|l_{00}|^2t}\rho_{00,0},\quad \rho_{10,t}=\rho_{01,t}^*,\\
  &&\rho_{11,t}=\rho_{11,0}+(1-e^{-|l_{00}|^2t})\rho_{00,0}, \\
  &&\rho_{01,t}=e^{-(\frac{1}{2}+\mbox{i})t}\rho_{01,0}+\int_0^te^{-(\frac{1}{2}+\mbox{i})(t-\tau)}e^{-|l_{00}|^2\tau}\rho_{00,0}d\tau.
  \end{eqnarray*}
Clearly, $\rho_t\to \rho'$ as $t\to\infty$; this implies that $\rho'$ is a
stable equilibrium of the Schr{\"o}dinger-picture evolution, as predicted by
Theorem~\ref{thl}. Finally, it is worth noting that this equilibrium state is a pure state (since
$\tr{{\rho'}^2}=1$). That is, the designed dissipation control does not lead to the decoherence of the
system.
\end{exmp}

In Example \ref{exam1}, the use of Condition DS also implies
the satisfaction of (\ref{gleq1}), and the system is exponentially
stable. However in general, Condition DS is a weaker
condition compared to Condition ES, as Condition DS does
not necessarily lead to exponential convergence. This can be illustrated in
the following example
\begin{exmp}\label{examdiff}\em
Consider a three-level system. Suppose we want to engineer the ground-state stability of the Lyapunov operator
\begin{equation}\label{exd1}
V=\left(\begin{array}{ccc}
0&0&0\\
0&1&0\\
0&0&2
\end{array}\right).
\end{equation}
To do this, the coupling operators are chosen as
\begin{equation}\label{exd2}
L_1=\left(\begin{array}{ccc}
0&1&0\\
0&0&0\\
0&0&0
\end{array}\right),\
L_2=\left(\begin{array}{ccc}
0&0&0\\
0&0&1\\
0&1&0
\end{array}\right).
\end{equation}
Using these values we compute
\begin{equation}\label{exd3}
\mathcal G(V)=\left(\begin{array}{ccc}
0&0&0\\
0&0&0\\
0&0&-1
\end{array}\right),\
\mathfrak{D}(V)=\left(\begin{array}{ccc}
0&0&0\\
0&2&0\\
0&0&1
\end{array}\right).
\end{equation}
Obviously, $\mathcal G(V)\leq0$. Additionally, we have
$\mathfrak{D}(V)\geq\frac{1}{2}V$. As a result, the Lyapunov operator
satisfies the sufficient condition DS. However, (\ref{gleq1}) does not hold
for $V$ and any $c>0$, hence it cannot be used to  establish exponential
stability. In this case, the dissipation will still steer the
system to the ground state, although the generator at the first-excited
state is zero. In fact, since the dissipation strength at the first-excited
state is two times the dissipation strength at the second-excited state,
the system will be partly driven to the ground state from the first-excited
state. However, there is also a possibility that the system will be
re-excited to the second-excited state, which makes the calculation of the
convergence speed difficult.
\end{exmp}
Example \ref{examdiff} provides the evidence that Condition ES
may not always follow from Condition DS. This example shows that asymptotic convergence is not always
exponential.

\section{Scalability of the Lyapunov methods}\label{secscal}
In this section, we address the question as to how the stability of
the subsystems scales up when these subsystems interact and
are coupled with  environments. Associated with the subsystems, consider a
collection of operators $\{W_\lambda\}$, $W_\lambda\succeq0$, $\lambda\in\Lambda=\{1,2,...,N\}$. Coupling between the system and
the environments is described by coupling operators $\{L_k\}$,
$k\in\Delta=\{1,2,...,K\}$. Also,
consider the operator in the
following form
\begin{equation}\label{sca1}
W=\sum_{\lambda=1}^N W_\lambda, \quad W_\lambda\succeq0,
\end{equation}
with the generator  and the dissipation functional of $W$ calculated to
be
\begin{eqnarray}\label{sca2}
\mathcal{G}(W)&=&\sum_{\lambda=1}^N\sum_k^KL_k^\dagger{W_\lambda}L_k-\frac{1}{2}L_k^\dagger{L_k}W_\lambda-\frac{1}{2}W_\lambda L_k^\dagger{L_k}\nonumber\\
&=&\frac{1}{2}\sum_{\lambda=1}^N\sum_k^K(L_k^\dagger[W_\lambda,L_k]+L_k^\dagger[W_\lambda,L_k]),
\\
\label{sca3}
\mathfrak{D}(W)&=&\sum_{\lambda^{'}=1}^N\sum_{\lambda=1}^N\sum_k^K[L_k^\dagger,W_\lambda][W_{\lambda^{'}},L_k].
\end{eqnarray}
In the space of the subsystems on which $W_\lambda$ associates, the generator and dissipation functional are
\begin{equation}\label{sca4}
\mathcal{G}(W_\lambda)=\sum_{k=1}^KL_k^\dagger{W_\lambda}L_k-\frac{1}{2}L_k^\dagger{L_k}W_\lambda-\frac{1}{2}W_\lambda L_k^\dagger{L_k},
\end{equation}
and
\begin{equation}\label{sca5}
\mathfrak{D}(W_\lambda)=\sum_{k=1}^K[L_k^\dagger,W_\lambda][W_\lambda,L_k],
\end{equation}
respectively. It follows from
(\ref{sca4})-(\ref{sca5}) that
\begin{equation}\label{newg1}
\mathcal{G}(W)=\sum_{\lambda=1}^N\mathcal{G}(W_\lambda),
\end{equation}
but in general
\begin{equation}\label{newd1}
\mathfrak{D}(W)\neq\sum_{\lambda=1}^N\mathfrak{D}(W_\lambda),
\end{equation}
which indicates that the dissipation behaviour may be quite different between $W$ and individual $W_\lambda$.

As noted, the scalability of stability relates to the preservation of stability under the operation of aggregation of stable subsystems. In
Subsection \ref{subsecfa} we consider the case where each subsystem possesses a Lyapunov observable that satisfies the conditions derived in the last
section. Precisely, we consider the situation where the coupling operators $\{L_k\}$ are such that each operator $W_\lambda$ satisfies
either Condition ES or Condition DS for this subsystem, i.e., each $W_\lambda$ is a Lyapunov operator and each subsystem, when considered
in isolation, asymptotically converges to the set of ground states of $W_\lambda$ . We then derive additional conditions which guarantee
that  $W=\sum_{\lambda=1}^N W_\lambda $ is a Lyapunov operator for the aggregated system, and hence the entire system is ground-state stable.

Another way to approach the scalability of the subsystems stability is via
studying the total system directly using Conditions ES and DS, without imposing the
ground-state stability requirement on individual subsystems and their
corresponding operators $W_\lambda$, which might be difficult if the
system is complex. This case is discussed in Subsection \ref{subsecsa}.

\subsection{Scalability of the ground-state stability of each $W_\lambda$}\label{subsecfa}
Combined with the results from Section \ref{ls}, the first statement in Lemma
\ref{ff} can be formulated in terms of the Lyapunov stability:
\begin{lem}\label{thmf1}
If Condition ES (respectively, Condition DS) holds for each $W_\lambda\succeq0$, then the system converges to the set of the ground states of $W$
  asymptotically. In addition, $W$ is a Lyapunov operator.
\end{lem}
\begin{pf*}{Proof. }
Lemma \ref{thmf1} directly follows from Lemma
  \ref{ff}.
\end{pf*}

The above lemma serves as the basis for the scalability analysis in this
section. The purpose of this analysis is to establish conditions for
preservation of the ground-state stability of $W$ when
aggregating ground-stable subsystems. To this end, suppose the ground
states of $W_\lambda$ are exponentially stabilized by coupling to the
$k$-th environment channel; the case of dissipative coupling will be
considered later. Specifically, suppose the coupling operator $L_k$ is such that
\begin{equation}\label{gl}
\mathcal{G}(W_\lambda)_{L_k}\leq-c_\lambda W_\lambda,\quad c_\lambda>0,
\end{equation}
where $\mathcal{G}(W_\lambda)_{L_k}$ denotes the single-channel component of the generator $\mathcal G(X)$,
\begin{equation}\label{oldgl}
\mathcal{G}(X)_{L_k}=L_k^\dagger{X}L_k-\frac{1}{2}L_k^\dagger{L_k}X-\frac{1}{2}XL_k^\dagger{L_k}.
\end{equation}

After the aggregation, the subsystems with which the observable $W_\lambda$ is associated, are subjected to other input
fields and so additional coherent couplings $\{L_{k^{'}},k^{'}\neq k,k\in\Delta\}$ are induced. Therefore, we need to ensure that these additional couplings
do not undermine the ground-state stability of $W_\lambda$. Formally, this
can be achieved by ensuring Condition ES is satisfied in the presence of
coupling with the environment channels other than $k$. For example, it is
sufficient to assume that
\begin{equation}\label{sd1}
\sum_{k^{'}=1,k^{'}\neq k}^K\mathcal{G}(W_\lambda)_{L_{k^{'}}}\leq0.
\end{equation}
Clearly, (\ref{sd1}) is a sufficient condition to guarantee that the
satisfaction of Condition ES can be established from (\ref{gl}). For this
reason, (\ref{sd1}) will be referred to as scalability condition. This
discussion is summarized in the following statement.

\begin{thm}\label{thmvl}
Suppose for each $\lambda\in\Lambda$, there exists an $L_k$ such that (\ref{gl}) and (\ref{sd1}) hold for each $W_\lambda$. Then, $W$ is asymptotically ground-state stable.
\end{thm}

The case where the ground states of $W_\lambda$ are asymptotically
stabilized using a dissipative coupling to the
$k$-th environment channel can be considered in the same manner. In this
case, the following statement holds.

\begin{thm}\label{thmv1new2}
Suppose for each $\lambda\in\Lambda$, there exists an $L_k$ such that
\begin{equation}\label{gl1}
\mathcal{G}(W_\lambda)_{L_k}\leq0,\quad \mathfrak{D}(W_\lambda)_{L_k}\geq
c_\lambda W_\lambda\ (\exists c_\lambda>0),
\end{equation}
and (\ref{sd1}) hold for each $W_\lambda$. Then, Condition DS is satisfied for $W$ and $W$ is asymptotically ground-state stable.
\end{thm}
\begin{pf*}{Proof. }
In the light of the previous discussion, we have $\mathcal{G}(W_\lambda)\leq0$ by (\ref{sd1}). Moreover, $\mathfrak{D}(W_\lambda)_{L_{k^{'}}}=[L_{k^{'}}^\dagger,W_\lambda][W_\lambda,L_{k^{'}}]$ is always non-negative for any $L_{k^{'}},k^{'}\neq k$, which yields the following relation
\begin{equation}\label{thmv1e1}
 \mathfrak{D}(W_\lambda)\geq\mathfrak{D}(W_\lambda)_{L_k}\geq c_\lambda
 W_\lambda.
\end{equation}
\hfill$\Box$
\end{pf*}
As Theorem \ref{thmv1new2} shows, we have dealt with the cross terms in
$\mathfrak{D}(W)$ by introducing a more conservative condition (\ref{sd1}),
which allowed us to engineer the condition (\ref{gl1}) on each dissipation
functional $\mathfrak{D}(W_\lambda)$ individually. More explicitly, by
stabilizing $W_\lambda$ separately and imposing the scalability condition
(\ref{sd1}), we can guarantee the convergence to the set of the ground
  states of $W$ without using the dissipation functional of the total
    system.

\subsection{Ground-state stability of $W$}\label{subsecsa}
As said before, the other approach to the scalability problem is to engineer
the ground-stability of the total system directly. One way to
achieve this is by induction, by grouping $W_\lambda$, $\lambda=1, \ldots,
n$, into $\widetilde{W}_n=\sum_{\lambda=1}^nW_\lambda$, and
considering $W_{n+1}$ as an additional observable. Note that we have
$n\in\{1,2,...,N\}=\Lambda$, and $W=\widetilde{W}_N$.
Then the algorithm to ensure $W$ is ground-state
stable is to iteratively achieve for each $n$ the ground-state stability of
$\widetilde{W}_{n+1}=\sum_{\lambda=1}^{n+1}W_\lambda=\widetilde{W}_n+W_{n+1}$,
by synthesizing coupling operators $\{L_k,\ k=M+1,\ldots,K\}$ additional to
the coupling operators $\{L_k,\ k=1,\ldots, M\}$ that ensure the
ground-state stability of $\widetilde{W}_n$.

Define $d_n$ as the smallest eigenvalue of $\widetilde{W}_n$, then we have
$\widetilde{W}_n-d_n\succeq0$. Obviously, $d_1=0$. The scalability
conditions arising from the above algorithm in the exponential and
asymptotic dissipation cases, respectively, are summarized in the following
theorems. The proof of these Theorems is given in the Appendix.

\begin{thm}\label{thmgeneral}
Suppose $\mathcal{G}(\widetilde{W}_n-d_n)\leq-c(\widetilde{W}_n-d_n)$,
$c>0$, is achieved using  a set of coupling operators $\{L_k,\ k=1,\ldots, M\}$. $\widetilde{W}_{n+1}$ is asymptotically ground-state stable if the additional coupling operators $\{L_k,\ k=M+1,\ldots,K\}$ satisfy the Lyapunov condition
\begin{eqnarray}\label{thmre1}
\mathcal{G}(W_{n+1})&+&\sum_{k=M+1}^K\mathcal{G}(\widetilde{W}_n)_{L_k}\nonumber\\
&\leq&-cW_{n+1}+c(d_{n+1}-d_{n}),\quad c>0.
\end{eqnarray}
\end{thm}
\begin{thm}\label{thmgeneralnew2}
Suppose the conditions
$\mathcal{G}(\widetilde{W}_n)\leq0,\mathfrak{D}(\widetilde{W}_n-d_n)\geq
c(\widetilde{W}_n-d_n)$, $c>0$ are achieved using a set of coupling
operators $\{L_k,\ k=1,\ldots, M\}$. The Lyapunov conditions to ensure the ground-state
stability of $\widetilde{W}_{n+1}$ are
\begin{eqnarray}
&&\mathcal{G}(W_{n+1})+\sum_{k=M+1}^K\mathcal{G}(\widetilde{W}_n)_{L_k}\leq0,\label{thmre2}\\
&&\mathfrak{D}(W_{n+1})+2\sum_{k=1}^K\mbox{Re}([L_{k}^\dagger,\widetilde{W}_n][W_{n+1},L_{k}])\nonumber\\
&&\geq cW_{\lambda={n+1}}-c(d_{n+1}-d_n),\ c>0.\label{thmre3}
\end{eqnarray}
Here $\{L_k,\ k=M+1,\ldots,K\}$ denote additional coupling operators.
\end{thm}
Interestingly, we can further obtain sufficient conditions for
Theorem \ref{thmgeneral} and Theorem \ref{thmgeneralnew2} without knowing
the value of $\{d_n\}$. Note that $W_{n+1}\succeq0$ and $d_n$ is the smallest eigenvalue of $\widetilde{W}_n$. So we
have
\begin{eqnarray}\label{drel}
&&d_{n+1}=\langle \widetilde{W}_{n+1}\rangle_{\rho_{g}^{n+1}}=\langle \widetilde{W}_n+W_{n+1}\rangle_{\rho_{g}^{n+1}}\nonumber\\
&&\geq \langle \widetilde{W}_n\rangle_{\rho_{g}^{n+1}}\geq d_n,
\end{eqnarray}
where $\rho_{g}^{n+1}$ denotes the ground state of
$\widetilde{W}_{n+1}$. Based on (\ref{drel}), we can obtain sufficient conditions which are not dependent on $\{d_n\}$.
\begin{cor}\label{coti}\mbox{}
\begin{enumerate}[(i)]
\item
Suppose the following condition holds:
\begin{eqnarray}
&&\mathcal{G}(W_{\lambda={n+1}})+\sum_{k=M+1}^K\mathcal{G}(\widetilde{W}_n)_{L_k}\leq-cW_{\lambda={n+1}},\nonumber\\
\label{rthmre1}
\end{eqnarray}
Then the conclusion of Theorem \ref{thmgeneral} holds.
\item
On the other hand, if
\begin{eqnarray}
&&\mathfrak{D}(W_{\lambda={n+1}})+2\sum_{k=1}^K\mbox{Re}([L_{k}^\dagger,\widetilde{W}_n][W_{\lambda={n+1}},L_{k}])\nonumber\\
&&\geq cW_{\lambda={n+1}},\ c>0,\label{rthmre3}
\end{eqnarray}
then the conclusion of Theorem \ref{thmgeneralnew2} holds.
\end{enumerate}
\end{cor}
\begin{pf*}{Proof. }
Using (\ref{drel}), it follows from condition (\ref{rthmre1})
(respectively, (\ref{rthmre3})) that (\ref{thmre1}) (respectively,
(\ref{thmre3})) holds. The statement of the Corollary then
follows from Theorem~\ref{thmgeneral} and \ref{thmgeneralnew2}.
\hfill$\Box$
\end{pf*}
If the additional coupling operators satisfy $\{[L_k,\widetilde{W}_n]=0,\ k=M+1,\ldots,K\}$, then the conditions of Corollary
\ref{coti} reduce to conditions (\ref{gl}) and (\ref{gl1}).

In contrast to the scalability approach considered in Lemma \ref{thmf1} and Theorem \ref{thmv1new2}, conditions (\ref{thmre3}) and
(\ref{rthmre3}) involve the cross-coupling terms
\begin{equation}
2\sum_{k=1}^K\mbox{Re}([L_{k}^\dagger,\widetilde{W}_n][W_{n+1},L_{k}]),
\end{equation}
and
\begin{equation}
2\sum_{k=1}^K\mbox{Re}([L_{k}^\dagger,\widetilde{W}_n][W_{\lambda={n+1}},L_{k}]).
\end{equation}
These cross-coupling terms show that
the condition on the dissipation functional of each $W_\lambda$ and the
condition on the dissipation functional of $W$ do not
necessarily imply each other. Therefore, the two methods to achieve
scalability proposed in this section have different implications.

We conclude this section with an illustration of the results of
Theorem \ref{thmgeneral} and the first statement of Corollary \ref{coti}.
\begin{exmp}\em
Recall the two-level system from Example~\ref{exam1}, which was concerned
with ground-state stabilization of the operator $W_1=\frac{1}{2}(1+\sigma_{z_1})=\left(\begin{array}{cc}
1&0\\
0&0
\end{array}\right)\succeq0$. The Pauli matrices are defined by
$\sigma_x=\left(\begin{array}{cc}
0&1\\
1&0
\end{array}\right),\sigma_y=\left(\begin{array}{cc}
0&-i\\
i&0
\end{array}\right), \sigma_z=\left(\begin{array}{cc}
1&0\\
0&-1
\end{array}\right)$. In example~\ref{exam1}, the coupling operator $L_1=\sigma_-=\left(\begin{array}{cc}
0&0\\
1&0
\end{array}\right)$ was found to satisfy the stability condition $\mathcal
G(W_1)_{L_1}\leq-W_1$. Now we consider the extended two-qubit system, on
which the operator of interests is $W=W_1+W_2$ with
$W_2=\frac{1}{2}(1+\sigma_{z_1}\sigma_{z_2})\succeq0$; here $W_1$
is the extended operator $W_1=W_1\otimes I_2$. Also, let us extend $L_1$ by
letting  $L_1=L_1\otimes I_2$. Condition
(\ref{rthmre1}) with $c=1$ in this case takes the form
\begin{equation}\label{examtwoe3}
\mathcal G(W_2)_{L_1}+\sum_{k=2}^3\mathcal G(W_2)_{L_k}+\sum_{k=2}^3\mathcal G(W_1)_{L_k}\leq-W_2,
\end{equation}
where $L_2$ and $L_3$ are the new coupling operators. Condition (\ref{examtwoe3}) can be further simplified
\begin{eqnarray}\label{examtwoe4}
\sum_{k=2}^3\mathcal G(W_2+W_1)_{L_k}&\leq&-W_2-\mathcal G(W_2)_{L_1}\nonumber\\
&=&-\left(\begin{array}{cccc}
0&0&0&0\\
0&1&0&0\\
0&0&0&0\\
0&0&0&1
\end{array}\right)=-\left(\begin{array}{cc}
0&0\\
0&W^{'}_{3\times3}
\end{array}\right).\nonumber\\
\end{eqnarray}
Since $W_1+W_2$ equals
\begin{equation}\label{examtwoe5}
W_1+W_2=\left(\begin{array}{cccc}
2&0&0&0\\
0&1&0&0\\
0&0&0&0\\
0&0&0&1
\end{array}\right)=\left(\begin{array}{cc}
2&0\\
0&W^{'}
\end{array}\right), \quad W^{'}=\left(\begin{array}{ccc}
1&0&0\\
0&0&0\\
0&0&1
\end{array}\right),
\end{equation}
letting $L_2,L_3$ have the form
\begin{equation}\label{examtwoe6}
L_2=\left(\begin{array}{cc}
1&0\\
0&L_{3\times3}^2
\end{array}\right),
L_3=\left(\begin{array}{cc}
1&0\\
0&L_{3\times3}^3
\end{array}\right)
\end{equation}
reduces inequality (\ref{examtwoe4}) to the inequality
\begin{equation}\label{examtwoe7}
\sum_{k=2}^3\mathcal G(W^{'})_{L_{3\times3}^k}\leq-W^{'},W^{'}\succeq0.
\end{equation}
The easiest way to solve (\ref{examtwoe7}) is
to further decompose $W^{'}$ as
\begin{equation}\label{examtwoe8}
W^{'}=\left(\begin{array}{ccc}
1&0&0\\
0&0&0\\
0&0&0
\end{array}\right)+\left(\begin{array}{ccc}
0&0&0\\
0&0&0\\
0&0&1
\end{array}\right),
\end{equation}
then the coupling operators are readily computed to be
\begin{equation}\label{examtwoe9}
L_2=\left(\begin{array}{cccc}
0&0&0&0\\
0&0&0&0\\
0&1&0&0\\
0&0&0&0
\end{array}\right),\ L_3=\left(\begin{array}{cccc}
0&0&0&0\\
0&0&0&0\\
0&0&0&1\\
0&0&0&0
\end{array}\right).
\end{equation}
\end{exmp}
In this example, since $W\succeq0$, the individual $W_{1,2}$ are also asymptotically ground-state stable. In addition, $W$ is frustration-free.

\section{Synthesis of the dissipation}\label{secapp}
In this section, we introduce the methods to find the correct dissipation controls that steer the system to the ground states of given candidate Lyapunov operators. Also, we will show how to calculate the system-environment couplings which satisfy the scalability conditions derived in Section \ref{secscal}.

\subsection{Synthesis in the case of single dissipation channel}
In this first part, we use a single candidate Lyapunov operator $V\succeq0$
and a single system-environment coupling operator $L$ as the dissipation
control. While we established previously that $L$ can be calculated using
Conditions ES and DS, these conditions generally lead to nonlinear matrix
inequalities. Solving these inequalities for large-scale systems is a
challenging task. In this section, we develop a method to circumvent these
difficulties.

We introduce a special class of dissipation controls
that admit factorization $L=UV$, where $U$ is a unitary operator. The
Reader who is interested in the physical realization of such an operator
$L$ can refer to the Appendix. In
\cite{Verstraete09} the authors have suggested similar coupling operators
$L_{i,\lambda}=U_iH_\lambda$ for the ground-state engineering of a
Hamiltonian $H_\lambda$. They showed that this class of control could form
a sufficient condition for ground-state stability if
$\{U_i\}$ is a set of unitary
operators which rotate part of the high-energy space with support in
$H_\lambda$ into the ground-state space, according to
\cite{Kraus08}. However, it is not clear when this rotation exists, and how
to solve for such unitary rotation.

In this section we characterize the unitary rotation $U$ required to establish the ground-state Lyapunov stability. Basically, we attempt to solve Conditions ES or DS for $U$. We have
\begin{eqnarray}\label{control1}
\mathcal G(V)&=&L^\dagger VL-\frac{1}{2}L^\dagger LV-\frac{1}{2}VL^\dagger L\nonumber\\
&=&VU^\dagger VUV-V^3
\end{eqnarray}
for single system-environment coupling $L$. We now consider several special choices for the operator $V$.

\subsubsection{Special case 1. $V$ is a projection ($V^2=V$)}
In many cases, $V$ can be constructed as a projection, i.e., $V^2=V$. For
example, the Hamiltonian discussed in \cite{Verstraete09,Perez08} can be
displaced by a constant to generate a Lyapunov operator $V$ which is also a
projection. Moreover, the two examples of physical relevance considered at the end of this section employ Lyapunov
operators which are projections.

With the aid of this property, Condition
ES can be rewritten as
\begin{equation}\label{cona1}
VU^\dagger VUV\leq(1-c)V,\ 0<c\leq1,
\end{equation}
which can be regarded as the mathematical formulation for the argument in
\cite{Verstraete09}: $U$ should be designed to rotate part of the
high-energy space into the zero-energy space. This shows that our stability
results are consistent with the physical intuition.

Now we turn to Condition DS. With $L=UV$, this
condition can be written as
\begin{equation}\label{cona6}
VU^\dagger VUV\leq V,
\end{equation}
and
\begin{eqnarray}\label{cona7}
\mathfrak{D}(V)=[VU^\dagger,V][V,UV]=-VU^\dagger VUV+V\geq cV,\\ (\exists
c>0). \nonumber
\end{eqnarray}
Obviously, (\ref{cona7}) implies (\ref{cona6}). More importantly, (\ref{cona7}) and (\ref{cona1}) are the same conditions. As a result, the sufficient conditions (\ref{gleq1}) and (\ref{gleq2}) for the ground-state stability of $V$ both reduce to the same expression (\ref{cona1}) under the assumptions $L=UV$ and $V^2=V$. In this case, Condition DS also leads to exponential convergence of $\langle V\rangle_{\rho_t}$ to $0$.

Letting $c=1$ in (\ref{cona1}) leads to a special case where $VUV=0$. In particular, a unitary rotation $U$ satisfying $VUV=0$ always exists when stable
states are engineered to be the ground states of $V$
\cite{Gottesman97,Verstraete09}.

In addition to the above special case, (\ref{cona1}) can be solved by
making the substitution $P=VUV$ which leads to the condition
\begin{equation}\label{cona2}
(1-c)V-P^\dagger P\geq0,\ 0<c<1.
\end{equation}
Since $V\succeq0$, it can be decomposed as $V=Q^\dagger Q$. Therefore, $P=\sqrt{1-c}Q$ is a solution to (\ref{cona2}). The synthesis problem is transformed to solving
\begin{equation}\label{cona8}
VUV=\sqrt{1-c}Q,\ 0<c\leq1
\end{equation}
for a unitary $U$. (\ref{cona8}) is equivalent to
\begin{equation}\label{cona9}
(V^T\otimes V)\mbox{vec}(U)=\mbox{vec}(\sqrt{1-c}Q),
\end{equation}
$\mbox{vec}(U)$ is the vectorization of an $n\times n$ matrix $U$ by
stacking the columns of $U$ into a single column vector of dimension $n^2\times 1$. The general solution to (\ref{cona9}) is given by
\begin{eqnarray}\label{cona10}
\mbox{vec}(U)&=&(V^T\otimes V)^+\mbox{vec}(\sqrt{1-c}Q)\nonumber\\
&+&(I_{n^2\times n^2}-(V^T\otimes V)^+(V^T\otimes V))x,
\end{eqnarray}
where $x$ is an $n^2\times 1$ vector of free parameters. $(V^T\otimes
V)^+$ denotes the unique Moore-Penrose pseudoinverse \cite{Lay02} of
$V^T\otimes V$. For convenience, we adopt the notations
\begin{eqnarray}
I_{n^2\times n^2}-(V^T\otimes V)^+(V^T\otimes V)&=&(a_1^T\quad a_2^T\quad \dots\quad a_n^T)^T,\nonumber\\
(V^T\otimes V)^+\mbox{vec}(\sqrt{1-c}Q)&=&(b_1^T\quad b_2^T\quad \dots\quad b_n^T)^T,\nonumber\\
\end{eqnarray}
where the elements $\{a_i,i=1,\dots,n\}$ are $n\times n^2$ matrices, and $\{b_i,i=1,\dots,n\}$ are $n\times 1$ vectors. According to (\ref{cona10}), $U$ can be expressed as
\begin{equation}\label{cona11}
U=(a_1x+b_1\quad a_2x+b_2\quad\dots\quad a_nx+b_n),
\end{equation}
which is an $n\times n$ matrix. The parameters $\{a_i\}$ and $\{b_i\}$ are
already known because $V,Q$ are given, and $x$ is determined from the
condition $U^\dagger U=I$. The latter condition can be explicitly written as
\begin{equation}\label{cona12}
\left(\begin{array}{c}
(a_1x+b_1)^\dagger\\
(a_1x+b_2)^\dagger\\
\vdots\\
(a_nx+b_n)^\dagger
\end{array}\right)(a_1x+b_1\quad a_2x+b_2\quad\dots\quad a_nx+b_n)=I.
\end{equation}
Equation (\ref{cona12}) can be further organized as a set of bilinear equations:
\begin{eqnarray}\label{cona13}
&&x^\dagger a_i^\dagger a_ix+x^\dagger a_i^\dagger b_i+b_i^\dagger a_ix+b_i^\dagger b_i=1,\ i=0,1,\dots,n,\nonumber\\
&&x^\dagger a_i^\dagger a_jx+x^\dagger a_i^\dagger b_j+b_i^\dagger a_jx+b_i^\dagger b_j=0,\ i>j.
\end{eqnarray}

The special case where $c=1$ and $VUV=0$ corresponds to
$b_1=b_2=\dots=b_n=0$. In this case, (\ref{cona13}) can be simplified as
\begin{eqnarray}\label{cona15}
x^\dagger a_i^\dagger a_ix&=&1,\ i=1,\dots,n,\quad x^\dagger a_i^\dagger a_jx=0,\ i>j.
\end{eqnarray}

\begin{exmp}\label{exam2}\em
For the purpose of illustrating the difference between the cases $c=1$ and
$c<1$, we again consider a quantum two-level system of
Example~\ref{exam1} where we considered the problem
of engineering the ground-state stability of the Lyapunov operator
(\ref{exame10}).
With $V$ defined in (\ref{exame10}), the Moore-Penrose pseudoinverse of
$V^{T}\otimes V$ is calculated to be
\begin{equation}\label{exam2e2}
(V^{T}\otimes V)^{+}=\left(\begin{array}{cccc}
1&0&0&0\\
0&0&0&0\\
0&0&0&0\\
0&0&0&0
\end{array}\right).
\end{equation}
First we solve (\ref{cona15}) in the case where $c=1$. $\{a_i\}$ can be obtained using (\ref{exam2e2}):
\begin{equation}\label{exam2e3}
a_1=\left(\begin{array}{cccc}
0&0&0&0\\
0&1&0&0
\end{array}\right),\ a_2=\left(\begin{array}{cccc}
0&0&1&0\\
0&0&0&1
\end{array}\right).
\end{equation}
With this, (\ref{cona15}) can be written as
\begin{eqnarray}\label{exam2e4}
&x^\dagger\left(\begin{array}{cccc}
0&0&0&0\\
0&1&0&0\\
0&0&0&0\\
0&0&0&0
\end{array}\right)x=1,\ x^\dagger\left(\begin{array}{cccc}
0&0&0&0\\
0&0&0&0\\
0&0&1&0\\
0&0&0&1
\end{array}\right)x=1,&\nonumber\\
&x^\dagger\left(\begin{array}{cccc}
0&0&0&0\\
0&0&0&0\\
0&0&0&0\\
0&1&0&0
\end{array}\right)x=0.&
\end{eqnarray}
Parameterizing $x$ as $x=(x_1\quad x_2\quad x_3\quad x_4)^{T}$, we arrive at a set of bilinear equations
\begin{eqnarray}\label{exam2e5}
&&|x_2|^2=1,\quad|x_3|^2+|x_4|^2=1,\quad x_4^*x_2=0.
\end{eqnarray}
Particularly, we have $x_4=0$ by (\ref{exam2e5}). Then the unitary rotation is
\begin{equation}\label{exam2e6}
U=(a_1x\quad a_2x)=\left(\begin{array}{cc}
0&x_3\\
x_2&0
\end{array}\right),\ |x_2|^2=|x_3|^2=1,
\end{equation}
and the desired system-environment coupling $L=UV$ is
\begin{equation}\label{exam2e7}
L=\left(\begin{array}{cc}
0&0\\
x_2&0
\end{array}\right).
\end{equation}
Next we consider the case when $c<1$. The decomposition $V=Q^\dagger Q$ is not unique, however due to the particular form of (\ref{exam2e2}), $(V^T\otimes V)^+\mbox{vec}(\sqrt{1-c}Q)$ is nonzero only if the first entry of $Q$ is nonzero. For example, we can choose $Q$ as the square root of $V$: $Q=\sqrt{V}=\left(\begin{array}{cc}
1&0\\
0&0
\end{array}\right)$, which gives
\begin{equation}\label{exam2e9}
b_1=\left(\begin{array}{c}
\sqrt{1-c}\\
0
\end{array}\right),\ b_2=\left(\begin{array}{c}
0\\
0
\end{array}\right).
\end{equation}
Equation (\ref{cona13}) transforms to
\begin{eqnarray}\label{exam2e10}
&&|x_2|^2=c,\quad|x_3|^2+|x_4|^2=1,\nonumber\\
&&x_4^*x_2+\sqrt{1-c}x_3^*=0,\quad m>0.
\end{eqnarray}
Accordingly, the unitary rotation and coupling are
\begin{eqnarray}
&U=\left(\begin{array}{cc}
\sqrt{1-c}&x_3\\
x_2&x_4
\end{array}\right),&\label{exam2e11}\\
&L=\left(\begin{array}{cc}
\sqrt{1-c}&0\\
x_2&0
\end{array}\right),\ |x_2|^2=c.&\label{exam2eadd}
\end{eqnarray}
Equation (\ref{exam2eadd}) gives the general form of the coupling operator $L$ which satisfies (\ref{cona1}) for
  $c<1$. Obviously, (\ref{exam2eadd}) does not incorporate the special case (\ref{exam2e7}) for $c=1$ since $|x_2|<1$.
\end{exmp}

\subsubsection{Special case 2. $V^2\geq V$}
In this case, the satisfaction of Conditions ES and DS still follows from
(\ref{cona1}) because
\begin{equation}\label{v21}
\mathcal G(V)=VU^\dagger VUV-V^3\leq VU^\dagger VUV-V\leq -cV.
\end{equation}
Additionally, we have
\begin{eqnarray}\label{cona16}
&&U^\dagger VU\leq U^\dagger V^2U, \\
\label{v22}
&&VU^\dagger VUV\leq VU^\dagger V^2UV.
\end{eqnarray}
Now it is easy to see that if $U$ satisfies (\ref{cona2}), then it also satisfies (\ref{cona1}), since
\begin{equation}\label{cona17}
VU^\dagger VUV\leq VU^\dagger V^2 UV\leq (1-c)V,\quad c>0.
\end{equation}
As a result, the unitary solution $U$ obtained from (\ref{cona2}) could work for both cases $V^2=V$ and $V^2\geq V$.

\subsection{Synthesis of multiple dissipation channels}
In this section we extend the coupling synthesis approach considered in the
previous section to construct multiple dissipation channels aimed at
ground-state stabilization of a Lyapunov observable $V$.
We still assume $V$ is a projection. Letting $L_k=U_kV$, we can re-express
(\ref{gleq1}) as
\begin{equation}\label{conb2}
\sum_{k=1}^KVU_k^\dagger VU_kV=V(\sum_{k=1}^KU_k^\dagger VU_k)V\leq(K-c)V,\ c>0.
\end{equation}
It is easy to verify that Condition DS is still equivalent
to Condition ES if we assume the decomposition $L_k=U_kV$ for each
$L_k$. This observation leads to the following ``no-go theorem'' concerning
the validity of such decomposition.
\begin{cor}
Suppose the Lyapunov observable $V$ is a projection.
If it satisfies Condition DS but Condition ES does not hold, then at least
one of the coupling operators $L_k$ does not admit decomposition of the
form $L_k=U_kV$ with a unitary $U_k$. In the single channel case, the
coupling operator $L$ cannot be represented as
form $L=UV$ with a unitary operator $U$.
\end{cor}
To illustrate the above result, consider Example \ref{examdiff} where Condition DS is not equivalent to Condition ES. We conclude
that the matrix $L_2$ in Example \ref{examdiff} cannot be written as
$L_2=U_2V$ where  $U_2$ is a unitary operator. If this decomposition was
possible, we would have
\begin{equation}\label{imde}
\left(\begin{array}{ccc}
0&0&0\\
0&0&1\\
0&1&0
\end{array}\right)=\left(\begin{array}{cc}
U_{00}&U_{01}\\
U_{10}&U_{11}
\end{array}\right)\left(\begin{array}{ccc}
0&0&0\\
0&1&0\\
0&0&2
\end{array}\right),
\end{equation}
where $U_{00}$ is a scalar. It is easy to see from (\ref{imde}) that $U_{11}=\left(\begin{array}{cc}
0&\frac{1}{2}\\
1&0
\end{array}\right),U_{01}=[0\quad0]$, which implies that $U_2$ cannot be a unitary operator.

The set of unitary operators $\{U_k\}$ satisfying (\ref{conb2}) can be
calculated if a decomposition such as $(K-c)V=\sum_{k=1}^KQ_k^\dagger Q_k$ is available. Then $\{U_k\}$ are obtained by solving $VU_kV=Q_k$, as did in (\ref{cona8}).

\subsection{Scalable dissipations}
In this section we are concerned with a particular class of coupling
operators of the form $L_k=U_kW_\lambda$, where $W_\lambda\succeq0,\lambda=1,2,...,N$ and $U_k$ is
a unitary operator. In this case, there is one-to-one correspondence between each $L_k$ and $W_\lambda$ and so we have $k=1,2,...,N$. For this type of coupling operators, we can re-express
the condition $[L_{k^{'}},W_\lambda]=0,k^{'}\neq k$, which is one
particular sufficient condition to guarantee satisfaction of the
scalability condition (\ref{sd1}), in terms of unitary operators
$U_k$. This leads to a sufficient condition for ground-state stability of
the operator $W=\sum_{\lambda=1}^N W_\lambda$, which follows from Theorems
\ref{thmvl} and \ref{thmv1new2}:
\begin{cor}\label{coro1}
Assume $[W_\lambda,W_{\lambda^{'}}]=0$ for $\lambda\neq\lambda^{'}$ and $[U_{k^{'}},W_\lambda]=0$ . If either (\ref{gl}) or (\ref{gl1}) holds for each $W_\lambda\succeq0$, then $W=\sum_{\lambda=1}^N W_\lambda$ is asymptotically ground-state stable.
\end{cor}
\begin{pf*}{Proof. }
The conclusion follows from
\begin{eqnarray}\label{coro1e2}
[L_{k^{'}},W_\lambda]&=&[U_{k^{'}},W_\lambda]W_{\lambda^{'}}=0,\ \lambda^{'}\neq\lambda.
\end{eqnarray}
Thus $\mathcal{G}(W_\lambda)_{L_{k^{'}}}=0$ and so the scalable condition (\ref{sd1}) used in Theorems
\ref{thmvl} and \ref{thmv1new2} is satisfied.\hfill$\Box$
\end{pf*}

If $W_\lambda$ is a projection, the general sufficient condition (\ref{sd1}) is then expressed as
\begin{equation}\label{gcs}
W_\lambda U_{k^{'}}^\dagger W_\lambda U_{k^{'}}W_\lambda\leq W_\lambda.
\end{equation}

We now present two examples of application of the scalable condition.
\begin{exmp}\label{examcs}\em
Consider the generation of one-dimensional cluster state for one-way quantum computation \cite{Rau01}. The system is composed of a chain of $N$ qubits with nearest neighbor interaction. The one-dimensional cluster state is the ground state of the candidate Lyapunov operator defined as
\begin{equation}\label{exam3e1}
W=\sum_{\lambda=2}^{N-1} W_\lambda,\quad W_\lambda=\frac{1}{2}(\sigma_{z_{\lambda-1}}\sigma_{x_\lambda}\sigma_{z_{\lambda+1}}+1).
\end{equation}
$\{W_\lambda\}$ are commuting due to
\begin{eqnarray}\label{exam3e2}
[W_\lambda,W_{\lambda+1}]&=&\frac{1}{4}\sigma_{z_{\lambda-1}}[\sigma_{x_\lambda}\sigma_{z_{\lambda+1}},\sigma_{z_\lambda}\sigma_{x_{\lambda+1}}]\sigma_{z_{\lambda+2}}=0.\nonumber\\
\end{eqnarray}
$U_\lambda=\sigma_{z_\lambda}$ is a solution to $W_\lambda U_\lambda W_\lambda=0$ and (\ref{gl}). Furthermore, we have
\begin{equation}\label{exam3e3}
[U_{\lambda^{'}},W_\lambda]=\frac{1}{2}[\sigma_{z_{\lambda^{'}}},\sigma_{z_{\lambda-1}}\sigma_{x_\lambda}\sigma_{z_{\lambda+1}}]=0
\end{equation}
for $\lambda \neq {\lambda^{'}}$. By Corollary \ref{coro1}, $W=\sum_{\lambda=2}^{N-1}
W_\lambda$ is asymptotically ground-state stable. In particular, $W$ can be
stabilized to its ground states by selecting the following operators as
coupling operators $\{L_\lambda=\sigma_{z_\lambda}(\sigma_{z_{\lambda-1}}\sigma_{x_\lambda}\sigma_{z_{\lambda+1}}+1)\}$.
\end{exmp}

\begin{figure}
\centering
\includegraphics[scale=0.6]{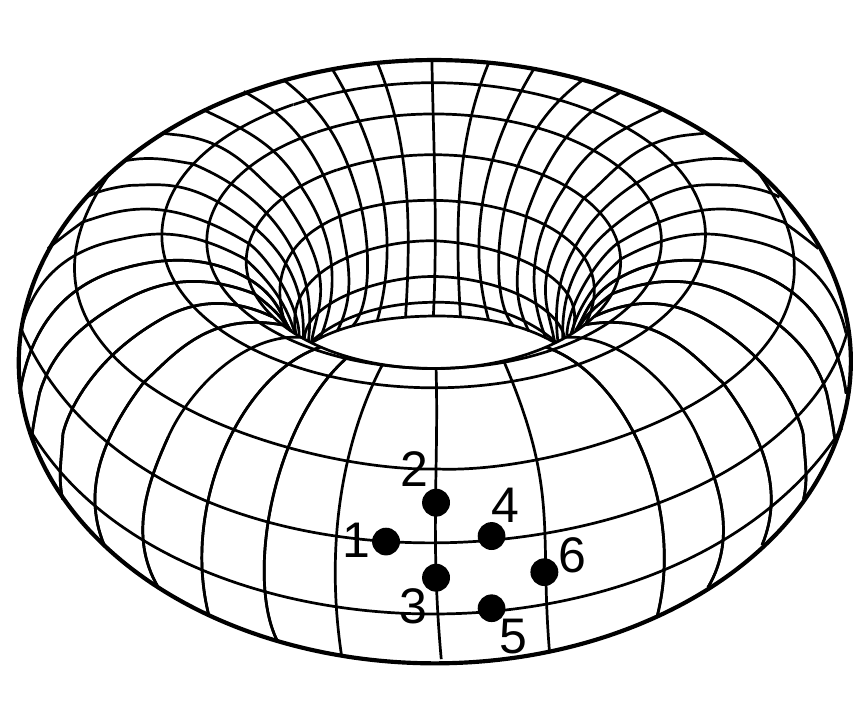}
\caption{The schematic representation of lattices on a torus with periodic boundary. The qubits are placed on the edges. The qubits $1,2,3,4$ are connected to a common vertex, and the qubits $3,4,5,6$ form a plaquette.}.
\label{figtorus}
\end{figure}

\begin{exmp}\label{exam4}\em
The toric code \cite{Kitaev03,Nielsen04,Eliot14} is defined on spin lattices. The qubits are placed on the edges, as shown in Figure \ref{figtorus}. The toric code states can then be defined by the degenerate ground states of the stabilizer operators as
\begin{equation}\label{exam4e1}
A=\prod_{i=1}^4\sigma_{x_i},\quad B=\prod_{j=3}^6\sigma_{z_j}.
\end{equation}
The code states are the ground states of both $A$ and
$B$. $\{\sigma_{x_i}\}$ are four X-axis Pauli operators acting on the four
qubits that connect to one vertex, and $\{\sigma_{z_j}\}$ are four Pauli
operators acting on the four qubits that form one plaquette. Errors can be easily detected and corrected using these code states as the computation basis. Normally, the code states are defined using a large number of stabilizer operators, and so the qubits could cover a large area of the torus.

First, we consider two stabilizer operators as outlined in (\ref{exam4e1}) and define the Lyapunov operator $V$ as $V=V_1+V_2,V_1=-\frac{1}{2}\sigma_{x_1}\sigma_{x_2}\sigma_{x_3}\sigma_{x_4}+\frac{1}{2},V_2=-\frac{1}{2}\sigma_{z_3}\sigma_{z_4}\sigma_{z_5}\sigma_{z_6}+\frac{1}{2}$. Performing the similar analysis as in the last example, any of the four operators $U_i=\sigma_{z_i},i\in\{1,2,3,4\}$ can be shown to stabilize $V_1$. Particularly, $\sigma_{z_i},i\in\{1,2,3,4\}$ commutes with any of the four operators $\{\sigma_{z_i},i=3,4,5,6\}$. In other words, $\sigma_{z_i},i\in\{1,2,3,4\}$ stabilizes $V_1$ without interfering with $V_2$. By Corollary \ref{coro1}, $V_1$ and $V_2$ are scalable and $V$ is asymptotically ground-state stable.

Suppose the code states are defined by the common ground states of three stabilizer operators, namely, $A$, $B$ and a third stabilizer operator as $V_3=-\frac{1}{2}\sigma_{x_1}\sigma_{x_7}\sigma_{x_8}\sigma_{x_9}+\frac{1}{2}$. $V_1$ and $V_3$ have one common edge. In Figure \ref{figtorus}, the qubits $7,8,9$ should be placed on the three edges connecting to the vertex on the left of qubit $1$. If we use $U=\sigma_{z_1}$ to stabilize $V_1$, then we have $[\sigma_{z_1},V_3]=[\sigma_{z_1},-\frac{1}{2}\sigma_{x_1}\sigma_{x_7}\sigma_{x_8}\sigma_{x_9}]\neq0$. The scalable condition in Corollary \ref{coro1} does not hold and so $\sigma_{z_1}$ acts non-trivially on $V_3$. However, it can be easily seen that $\sigma_{z_1}$ indeed stabilizes $V_3$ as well, which can be considered as a special case where (\ref{sd1}) holds.
\end{exmp}

\section{Conclusion}\label{conc}
We have developed the Lyapunov theory of the ground-state stability of
quantum systems using a Heisenberg-picture approach. This theory is
designed to serve as a foundation for a theory of stabilization by
dissipation, which has significant applications in future quantum
technologies. This theory allows us to engineer the systems by considering
Lyapunov operators and manipulating inequalities involving the system
generators applied to these operators, which is a common practice when
engineering classical and quantum control systems. Several issues should be taken into consideration concerning practical implementations of this theory. For example, the realization of the resulting coupling operators $\{L_k\}$ with the available experimental resources is an important and interesting problem. Additional Hamiltonian control could be introduced if the choices of dissipation channels are limited, as was done in \cite{Ticozzi09,Ticozzi12}. The convergence speed is also critical for a large-scale quantum system. We have shown that the dissipative stability condition DS does not necessarily guarantee exponential convergence. Therefore, the scaling of the convergence speed as we build up the systems using weak Lyapunov conditions will need further investigation.





\section{Appendix}\label{appa}

\subsection{Proof of Proposition \ref{df3}}

If $X$ is asymptotically ground-state stable, then $\langle X\rangle_{\rho_t}\rightarrow d$ by Definition \ref{defgs}. Since $\rho_t$ is tight, the limit point always exists. For any limit point $\rho^{'}$ which is the limit of a converging subsequence $\rho_{t_k}$
satisfying $\rho_{t_k}\rightarrow\rho^{'}$, we have $\langle X\rangle_{\rho^{'}}=\lim_{k\rightarrow\infty}\langle
X\rangle_{\rho_{t_k}}=\lim_{t\rightarrow\infty}\langle X\rangle_{\rho_t}=d$ and so $\rho^{'}$ is contained in $Z_X$. If $\rho_t$ is converging to $Z_X$, then for any converging subsequence $\rho_{t_k}$ of $\rho_t$ we have $\langle X\rangle_{\rho_{t_k}}\rightarrow\langle X\rangle_{\rho^{'}}=d$, where $\rho^{'}$ is the limit point. Therefore, $\lim_{t\rightarrow\infty}\langle X\rangle_{\rho_t}$ exists and equals $d$.

\subsection{Proof of Lemma \ref{ff}}

To prove (i) we observe that if $\langle
W_\lambda\rangle_{\rho_t}\rightarrow0$, then $\langle
W\rangle_{\rho_t}=\sum_\lambda\langle
W_\lambda\rangle_{\rho_t}\rightarrow0$. It remains to show that $d=0$. Since $W$
is a finite-dimensional
operator, it has finite number of eigenvalues. Suppose the smallest
eigenvalue of $W$ is positive, i.e., $d>0$. Then $W- d\ge 0$ and $\langle
(W-d)\rangle_{\rho_t}\ge 0$  for any state $\rho_t$. Thus, $\langle
W\rangle_{\rho_t}\ge d>0$ for any state $\rho_t$ (since
$\tr{\rho_t}=1$). This contradicts $\langle W\rangle_{\rho_t}\rightarrow0$. Conversely, if $\langle W\rangle_{\rho_t}\rightarrow0$, then $\langle
W_\lambda\rangle_{\rho_t}\rightarrow0$ as $\langle
W_\lambda\rangle_{\rho_t}\succeq0$ for each $\lambda$, which proves the ground-state stability of each $W_\lambda$.

\subsection{Proof of Lemma \ref{theorem1}}

Since $\mathcal{G}(V)\leq0$, $\langle
V\rangle_{\rho_t}\le \langle V\rangle_{\rho_0}$ and $\lim_{t\rightarrow\infty}\langle
V\rangle_{\rho_t}$ exists. Recall that since $V$ is a Lyapunov operator,
then $Z_V=\{\rho: \langle V\rangle_\rho=0\}$. Hence $\forall \rho_0\in
Z_V$, $\langle V\rangle_{\rho_t}= \langle V\rangle_{\rho_0}=0$; this implies
that $Z_V$ is an invariant set. We only need to prove that
$\rho_t$ will exit the domain $\{\rho:\langle V\rangle_\rho\geq\epsilon\}$
for arbitrary $\epsilon>0$. Now suppose the trajectory $\rho_t$ is
restricted to a domain $\{\rho:\langle V\rangle_\rho\geq\epsilon\}$ for
some $\epsilon>0$. There exists an
invariant state $\rho_I$ which is the limit point of the tight sequence $\frac{1}{t}\int_0^{t}\rho_{t^{'}}dt^{'}$ \cite{Pan14}. $\frac{1}{t}\int_0^{t}\rho_{t^{'}}dt^{'}$ is the mean of
$\rho_t$, and so $\rho_I$ is in $\{\rho:\langle
V\rangle_\rho\geq\epsilon\}$. By assumption,
$\langle\mathcal{G}(V)\rangle_{\rho_I}<0$. Let the initial state be the invariant state $\rho_I$.
Integrating $\mathcal{G}(V)$ over $[0,t]$ yields
\begin{equation}\label{e1ly1}
\langle V\rangle_{\rho_I}-{\langle V\rangle_{\rho_I}}=0=\int_0^t\langle \mathcal G(V)\rangle_{\rho_I}dt^{'}.
\end{equation}
This leads to a contradiction as
$\langle\mathcal{G}(V)\rangle_{\rho_I}<0$. We thus conclude that for any
$\epsilon>0$, there exists $t(\epsilon)$ such that  $\langle
V\rangle_{\rho_t}\le \langle V\rangle_{\rho_{t(\epsilon)}}<\epsilon$
  $\forall t>t(\epsilon)$. That is, $V$ is asymptotically ground-state stable.

\subsection{Proof of Lemma \ref{theorem4}}

Similar to the proof of Lemma \ref{theorem1} and the proof of Proposition
\ref{theorem3} in \cite{Pan14}, we only need to prove that  for arbitrary
$\epsilon>0$,  the domain $\{\rho:\langle
V\rangle_\rho\geq\epsilon\}$ does not contain invariant states
$\rho_I$. Suppose this is not true and there is an invariant state $\rho_I$
in the domain $\{\rho:\langle V\rangle_\rho\geq\epsilon\}$. Consider the
positive operator $W=V^2$. The generator for $W$ is
\begin{equation}\label{e1t4}
\mathcal{G}(W)=V\mathcal{G}(V)+\mathcal{G}(V)V+\mathfrak{D}(V).
\end{equation}
Let the initial state be the invariant state $\rho_I$. Integrating $\mathcal{G}(W)$ leads to
\begin{eqnarray}\label{e3t4}
\langle W\rangle_{\rho_I}-{\langle W\rangle_{\rho_I}}&=&\int_0^t\langle V\mathcal{G}(V)+\mathcal{G}(V)V+\mathfrak{D}(V)\rangle_{\rho_I}dt^{'}\nonumber\\
&=&0.
\end{eqnarray}
To establish a contradiction, we use the following identity
\begin{equation}\label{e2t4}
\frac{1}{c}V\mathcal{G}(V)+\frac{1}{c}\mathcal{G}(V)V+V^2+\frac{1}{c^2}\mathcal{G}(V)^2=[V+\frac{1}{c}\mathcal{G}(V)]^2.
\end{equation}
Note that by assumption, $\langle
  V+\frac{1}{c}\mathcal{G}(V)\rangle_{\rho_I}=\langle
  V\rangle_{\rho_I}\ge \epsilon$, and hence
$\langle[V+\frac{1}{c}\mathcal{G}(V)]^2\rangle_{\rho_I}\ge \epsilon^2
>0$ due to the
positivity of the variance $\langle X^2\rangle_\rho-\langle
X\rangle^2_\rho\geq0$ for any Hermitian operator $X$. Using
$\langle[V+\frac{1}{c}\mathcal{G}(V)]^2\rangle_{\rho_I}>0$ we have
\begin{eqnarray}\label{e4t4}
&&\langle V\mathcal{G}(V)+\mathcal{G}(V)V+\mathfrak{D}(V)\rangle_{\rho_I}\nonumber\\
&>&\langle-cV^2-\frac{1}{c}\mathcal{G}(V)^2+\mathfrak{D}(V)\rangle_{\rho_I}.
\end{eqnarray}
Next, choose a positive number $d>0$ such that $\mathcal G(V)+d\geq0$, and
then we have $-\mathcal G(V)(\mathcal G(V)+d)\geq0$  since $-\mathcal G(V)$
and $\mathcal G(V)+d$ commute. The latter inequality can be
written as $-\mathcal{G}(V)^2\geq d\mathcal G(V)$. This results in the
following inequality
\begin{eqnarray}\label{e5t4}
\lefteqn{\langle
  V\mathcal{G}(V)+\mathcal{G}(V)V+\mathfrak{D}(V)\rangle_{\rho_I}} &&
\nonumber\\
&>&\langle-cV^2-\frac{1}{c}\mathcal{G}(V)^2+\mathfrak{D}(V)\rangle_{\rho_I}\nonumber\\
&\geq&\langle-cV^2+\frac{d}{c}\mathcal{G}(V)+\mathfrak{D}(V)\rangle_{\rho_I}\nonumber\\
&=&\langle-cV^2+\mathfrak{D}(V)\rangle_{\rho_I}\nonumber\\
&\geq&0.
\end{eqnarray}
The last line of (\ref{e5t4}) is obtained using the assumption
$cV^2\leq\mathfrak{D}(V)$. As a consequence, (\ref{e3t4}) is not
consistent with (\ref{e5t4}). This
contradiction shows that   for arbitrary
$\epsilon>0$,  the domain $\{\rho:\langle
V\rangle_\rho\geq\epsilon\}$ does not contain invariant states
$\rho_I$, hence any trajectory $\rho_t$ must exit the set $\{\rho:\langle
V\rangle_\rho\geq\epsilon\}$. This conclusion results in the asymptotic
ground-state stability of $V$, which can be established using the same
argument as in the proof of Lemma~\ref{theorem1}.

\subsection{Proof of Lemma \ref{theorem5}}

Choose a positive number $d>0$ such that $V-d\leq0$, from which we can
conclude $V(V-d)\leq0$ since $V$ and $(V-d)$ commute. This can be rewritten
as $dV\geq V^2$. Thus we have
\begin{equation}\label{t5e1}
V^2\leq dV\leq \frac{d}{c}\mathfrak{D}(V).
\end{equation}
By Lemma \ref{theorem4}, $\rho_t$ converges to the ground states of $V$.

\subsection{Proof of Theorem \ref{thmgeneral} and \ref{thmgeneralnew2}}

For simplicity, first we
consider the integration of two operators represented by $W_1$ and $W_2$
with $W_{1,2}\succeq0$. $W=W_1+W_2$ could be positive definite. Since our Lyapunov stability
results are derived under the assumption that the candidate Lyapunove
operator has zero eigenvalue, we circumvent this issue by considering the
displaced operator $W-d\succeq0$, where $d\geq0$ is the smallest eigenvalue
of $W$. Suppose the Lyapunov condition $\mathcal{G}(W_1)_{L_1}\leq-cW_1$
has been established using the coherent coupling $L_1$. We are concerned
with engineering an additional coupling between the environment and the
part of the system characterized by the observable $W_2$ to achieve
the following Lyapunov condition for the total system
\begin{eqnarray}\label{lyt1}
\mathcal{G}(W-d)&=&\mathcal{G}(W_1+W_2)\leq-c(W-d)\nonumber\\
&=&-c(W_1+W_2)+cd,\quad c>0.
\end{eqnarray}
Formally, this problem reduces to the that of the synthesis of
an additional coupling operator $L_2$ which couples an environment to $W_2$. Decomposing (\ref{lyt1}) yields
\begin{eqnarray}\label{lyt2}
\mathcal{G}(W_1)_{L_1}+\mathcal{G}(W_1)_{L_2}+\sum_{k=1,2}\mathcal{G}(W_2)_{L_k}&&\nonumber\\
\leq-c(W_1+W_2)+cd,&&\quad c>0.
\end{eqnarray}
Since $\mathcal{G}(W_1)_{L_1}\leq-cW_1$, a sufficient condition for (\ref{lyt2}) to hold is
\begin{equation}\label{lyt3}
\mathcal{G}(W_1)_{L_2}+\mathcal{G}(W_2)\leq-cW_2+cd.
\end{equation}
It follows from the above discussion that $W$ is asymptotically
ground-state stable if (\ref{lyt3}) holds. Similar results can be obtained based on the assumption
\begin{equation}\label{lyt4}
\mathcal{G}(W_1)_{L_1}\leq0,\quad \mathfrak{D}(W_1)_{L_1}\geq cW_1,\ c>0.
\end{equation}
In order to engineer the stability of the combined system achieving
\begin{eqnarray}\label{lyt5}
&&\mathcal{G}(W-d)=\mathcal{G}(W_1+W_2)\leq0,\nonumber\\
&&\mathfrak{D}(W-d)=\mathfrak{D}(W_1+W_2)\nonumber\\
&\geq&c(W-d)=c(W_1+W_2)-cd,\ c>0,
\end{eqnarray}
we exploit the following relation
\begin{eqnarray}\label{f1e1}
&&\mathfrak{D}(\sum_{\lambda\in\Lambda} W_\lambda)=\sum_{k\in\Delta;\lambda,\lambda^{'}\in\Lambda}[L_k^\dagger,W_\lambda][W_{\lambda^{'}},L_k]\nonumber\\
&=&\sum_{\lambda\in\Lambda}\mathfrak{D}(W_\lambda)+\sum_{k\in\Delta;\lambda\neq\lambda^{'}}[L_k^\dagger,W_\lambda][W_{\lambda^{'}},L_k],
\end{eqnarray}
to write the second inequality in (\ref{lyt5}) explicitly as
\begin{eqnarray}\label{lyt6}
&&\mathfrak{D}(W_1)+\mathfrak{D}(W_2)+\sum_{k,\lambda,\lambda^{'}=1,2;\lambda\neq\lambda^{'}}[L_k^\dagger,W_\lambda][W_{\lambda^{'}},L_k]\nonumber\\
&\geq&\mathfrak{D}(W_1)_{L_1}+\mathfrak{D}(W_2)+\sum_{k,\lambda,\lambda^{'}=1,2;\lambda\neq\lambda^{'}}[L_k^\dagger,W_\lambda][W_{\lambda^{'}},L_k]\nonumber\\
&\geq&c(W_1+W_2)-cd.
\end{eqnarray}
Accordingly, a sufficient condition to guarantee $\mathfrak{D}(W-d)\geq c(W-d)$ is
\begin{equation}\label{lyt7}
\mathfrak{D}(W_2)+\sum_{k,\lambda,\lambda^{'}=1,2;\lambda\neq\lambda^{'}}[L_k^\dagger,W_\lambda][W_{\lambda^{'}},L_k]\geq cX_2-cd.
\end{equation}
The above methods can be readily extended to consider a system that involves an
arbitrary number of subsystems and observables $W_\lambda$, resulting in Theorem \ref{thmgeneral} and \ref{thmgeneralnew2}.

\subsection{Physical implementation of the coupling operator $L$}
The key component for the proposed method is the ability to physically
realize  the engineered coupling $L$. In principle, arbitrary dynamical
open quantum system can be implemented with high degree of precision given
the appropriate parameter scaling in the following sense
\begin{equation}
\lim_{k\to\infty} \sup_{0\leq t \leq T}|| U^{(k)}(t)\psi - U(t)\psi|| = 0, \qquad \forall \psi \in \mathcal{H},
\end{equation}
in which $\mathcal{H}$ is a subspace of the total Hilbert space. The  $U^{(k)}(t)$ and $U(t)$ are the pre-limit and limit unitary operators given by
\begin{eqnarray}
dU^{(k)}(t)&=&(-iH^{(k)}dt + dB^\dagger(t)\tilde{L}^{(k)}-\tilde{L}^{(k)\dagger}dB(t)\nonumber\\
&-&\frac{1}{2}\tilde{L}^{(k)\dagger}\tilde{L}^{(k)}dt)U^{(k)}(t),  \qquad U^{(k)}(0) = I,\nonumber\\
dU(t)&=&( -iHdt + dB^\dagger(t)\tilde{L}-\tilde{L}^{\dagger}dB(t)\nonumber\\
&-&\frac{1}{2}\tilde{L}^{\dagger}\tilde{L}dt)U(t), \qquad U(0) = I.
\end{eqnarray}
The implementation of $L$ is done by adding an ancillary qubit to the
principal system. The principal system and the ancillary qubit are
subjected to the following Hamiltonian
\begin{equation}
H^{(k)} = k \Omega (L\sigma_++L^\dag\sigma_-)+H_S,
\label{eq: bilinear_interaction_Hamiltonian}
\end{equation}
where $\sigma_+,\sigma_-$ are operators of the ancillary qubit. In
addition, the qubit is coupled to an environment via the coupling
$\tilde{L}^{(k)} = k\sqrt{\gamma}\sigma_-$, where $k\sqrt{\gamma}$ is the
decay rate of the ancillary qubit. In the limit of fast decay of the qubit
($k\rightarrow\infty$) the principal system defined on $\mathcal H$
can be approximated using $\tilde{L} = -\frac{\Omega}{\sqrt{\gamma}}L, \qquad H = H_S$. This is commonly known in quantum optics as adiabatic elimination, where a fast degree of freedom is eliminated. Thus, we have a general approach to synthesize the desired coupling operator $L$. This method has been suggested for dissipation engineering in \cite{Verstraete09,Nurdin2009}. A recent experimental implementation of this method can be found in \cite{Kienzler02012015}.

\end{document}